\documentclass{fundam} 

\publyear{25}
\papernumber{3}
\volume{194}
\issue{3}
\theDOI{10.46298/fi.11547}

\versionForARXIV

\usepackage{url}
\usepackage{alltt}
\usepackage{amsmath} 
\usepackage{amssymb}

\usepackage{epsfig}
\usepackage{wrapfig}
\usepackage{subfigure}

\usepackage{rotating}
\usepackage{moreverb}
\usepackage{fancyvrb}

\def\bd{\begin{description}}
\def\ed{\end{description}}
\def\bc{\begin{center}}
\def\ec{\end{center}}
\def\bq{\begin{quote}}
\def\eq{\end{quote}}
\def\bi{\begin{itemize}}
\def\ei{\end{itemize}}
\def\be{\begin{enumerate}}
\def\ee{\end{enumerate}}
\def\ba{\begin{array}}
\def\ea{\end{array}}

\newcommand{\qed}{\hfill \ensuremath{\Box}} 

\newcommand{\chr}{{CHR}}
\newcommand{\true}{{\mathit true}}

\newcommand \aeq {=}  
\newcommand{\false}{{\mathit false}}

\newcommand{\lv}[1]{ }  

\newcommand{\CT}{\ensuremath{\mathcal{CT}}}

\newcommand{\simp}{{\; \Leftrightarrow \;}}

\newcommand{\myparagraph}[1]{\textbf{#1.}}

\begin{document}

\title{%
Runtime Repeated Recursion Unfolding in CHR: A Just-In-Time Online Program Optimization Strategy That Can Achieve Super-Linear Speedup}  %

\address{Ulm University, 89069 Ulm, Germany}

\author{%
Thom Fr{\"u}hwirth\\
University of Ulm, Germany\\
thom.fruehwirth{@}uni-ulm.de
}

\runninghead{T.~Fr{\"u}hwirth}{Runtime Repeated Recursion Unfolding: Online Program Optimization for Super-Linear Speedup}

\maketitle

\vspace*{-2cm} %

\begin{abstract}

We introduce a just-in-time runtime program transformation strategy based on 
repeated recursion unfolding.
Our online program optimization generates several versions of a recursion differentiated by the minimal number of recursive steps covered.
The base case of the recursion is ignored in our technique.

Our method is introduced here on the basis of single linear direct recursive rules. 
When a recursive call is encountered at runtime, first an unfolder creates specializations of the associated recursive rule on-the-fly and then an interpreter applies these rules to the call.
Our approach reduces the number of recursive rule applications to its logarithm 
at the expense of introducing a logarithmic number of generic unfolded rules.

We prove correctness of our online optimization technique and determine its time complexity. 
For recursions which have enough simplifyable unfoldings, a super-linear is possible, 
i.e. speedup by more than a constant factor.
The necessary simplification is problem-specific and has to be provided at compile-time.
In our speedup analysis, we prove a sufficient condition as well as a sufficient and necessary condition for super-linear speedup relating the complexity of the recursive steps of the original rule and the unfolded rules.

We have implemented an unfolder and meta-interpreter for runtime repeated recursion unfolding with just five rules in Constraint Handling Rules (CHR) embedded in Prolog.
We illustrate the feasibility of our approach with simplifications, time complexity results and benchmarks for some basic tractable algorithms.
The simplifications require some insight and were derived manually.
The runtime improvement quickly reaches several orders of magnitude, consistent with the super-linear speedup predicted by our theorems.
\end{abstract}




\section{Introduction}\label{intro}

In the context of rule-based programming,
\emph{unfolding} is a program transformation that basically replaces a call in the body (right-hand side) of a rule with the body of a rule whose head (left-hand side) is matched by the call. 
\emph{Repeated recursion unfolding} \cite{reclopstr}
first unfolds a given recursive rule with itself and simplifies it. 
This results in a specialized recursive rule that covers two recursive steps instead of one.
It continues to unfold the last unfolded recursive rule with itself. 
Each unfolding doubles the number of recursive steps covered by the unfolded rule.
In this article, we extend the method to an online program optimization and give an implementation of the necessary unfolder and interpreter. 
The given call determines how far the unfolding proceeds and how many rules are generated.
Therefore the optimization cannot be performed at compile-time.

\begin{example}[Summation]
Consider the following toy example, a simple recursive program
written in abstract syntax of the programming language Constraint Handling Rules (CHR). It recursively adds all numbers from $1$ to $n$.
The rules can be understood as a procedure definition for a binary relation $sum$. 
Rule $b$ covers the base case and rule $r$ covers the recursive case.
\begin{gather*}
  b: sum(N,S) \simp N=1 \, |\, S=1\\
  r: sum(N,S) \simp N>1 \, |\, sum(N{-}1,S1), S \aeq N{+}S1
\end{gather*}
The head $sum(N,S)$, guard (e.g. $N=1$) and body of a rule are separated by the symbols $\simp$ and $|$, respectively.
Upper case letters stand for variables.
When a call matches the head of a rule and the guard condition holds, the body of the rule is executed.

Unfolding the recursive rule with a copy of itself and simplifying the resulting rule gives
$$r_1: sum(N,S) \simp N>2 \, |\, sum(N{-}2,S1'), S \aeq 2{*}N{-}1{+}S1'.$$
Note that this rule $r_1$ cannot replace the original recursive rule because it only applies in case $N>2$. It behaves like applying the original rule $r$ twice. 
With rule $r_1$ we only need about half as many recursive steps as with the original rule alone. 
Because the arithmetic computation is simplified, we can also expect to halve the runtime.

We can now unfold rule $r_1$ with itself:
$$r_2: sum(N,S) \simp N>4 \, |\, sum(N{-}4,S1), S \aeq 4*N{-}6+S1$$
This rule results in fourfold speedup.
We can continue this process, doubling the speed each time\footnote{Clearly there is a closed form solution for this problem, $S=N*(N+1)/2$, but this is not the point of the example.}.

The most unfolded rule should cover as many recursive steps of the call as possible but not more.
For example, 
for $N{=}4$ we will unfold till rule $r_1$ with guard $N{>}2$,
for $N{=}5$ we will unfold till rule $r_2$ with $N{>}4$,
for $N{=}50$ we will unfold till rule $r_5$ with $N{>}32$.
\end{example}

As we have just seen,
our method requires unfolding on-the-fly because the number of unfoldings depends on the current call. 
We do not want to modify the given program at runtime.
Therefore we also introduce a simple interpreter for the unfolded rules.
This \emph{meta-interpreter\footnote{A meta-interpreter interprets a program written in its own implementation language.}} tries and applies each unfolded rule at most once starting with the given call and the most unfolded rule.
With sufficient simplification of the unfolded rules (as in the example), a super-linear speedup in runtime can be achieved. The time complexity is reduced.

\myparagraph{Overview and Contributions of the Paper}
In this paper, we introduce our online program optimization strategy of runtime repeated recursion unfolding as a systematic way to enable significant speedups.
We assume a single recursive rule with linear direct recursion and focus on tractable problems, i.e. those with polynomial worst-case time complexity.
We will use summation as our running example.

\emph{Section \ref{prelim}} 
recalls syntax and semantics of the CHR programming language.

\emph{Section \ref{rrrun}} defines our program transformation method of runtime repeated recursion unfolding with simplification and proves it correct.
We also show that there is an optimal rule application strategy and prove it sound and complete.

\emph{Section \ref{impl}} presents our lean implementation of the unfolder and meta-interpreter to perform repeated recursion unfolding at runtime. 

\emph{Section \ref{impltime}} derives the worst-case time complexity of our unfolder and meta-interpreter in relation to that of the given recursive rule using recurrence equations.

\emph{Section \ref{secspeedup}} proves a sufficient condition as well as a sufficient and necessary condition for super-linear speedup relating the complexity of the recursive steps of the original rule and the unfolded rules.

\emph{Section \ref{secbench}} contains the experimental evaluation of our technique on three examples, summation, list reversal and sorting. We derive the necessary simplifications, analyse their time complexity and compare it with the result of benchmarks.

\emph{Section \ref{related}}
discusses related work
and \emph{Section \ref{discuss}} discusses potential limitations and possible improvements of our approach. 
Finally, we end with conclusions and future work.

\section{Preliminaries}\label{prelim}

We recall the abstract syntax and the equivalence-based abstract operational semantics of CHR (Constraint Handling Rules) \cite{fru_chr_book_2009,rpbook} in this section. 

\subsection{Abstract Syntax of CHR}

The CHR language is based on the abstract concept of constraints.
{\em Constraints} are relations, distinguished predicates of first-order predicate logic.
There are two kinds of constraints: {\em built-ins (built-in constraints)} 
and
{\em user-defined (CHR) constraints} 
which are defined by the rules in a CHR program.
Built-ins can be used as tests in the guard as well as for auxiliary computations in the body of a rule.
There are at least the built-in constraints 
$\true$ and $\false$ (denoting inconsistency), 
including equality $=$ over terms with arithmetic expressions
and the usual relations over arithmetic expressions.
When CHR is embedded into a host language,
host language statements are regarded as built-ins.

\begin{definition}[CHR Program and Rules]
{
A {\em \chr\ program} is a finite set of rules.  
A {\em (generalized simplification) rule} is of the form
\[r: H \Leftrightarrow C \, |\, B,\]
where $r$ is an optional {\em name} (a unique identifier) of a rule.
The {\em head} $H$ is a conjunction of user-defined constraints,
the optional {\em guard} $C$ is a conjunction of built-ins,
and the {\em body} $B$ is a goal.
The {\em local variables of a rule} are those not occurring in the head of the rule.
A {\em goal} is a conjunction of built-in and user-defined constraints.
A {\em call} is either an atomic constraint in a rule body or a given constraint.
A {\em linear direct recursive rule} has exactly one call that has the same constraint symbol as the single head constraint.

} 
\end{definition}
(Possibly empty) conjunctions of constraints are denoted by upper case letters in definitions, lemmas and theorems.
Conjunctions are understood as multisets of their atomic conjuncts.
To avoid clutter, we often use simple commas to denote logical conjunction.

\subsection{Abstract Operational Semantics of CHR}\label{sec:chr:semantics}

Computations in CHR are sequences of rule applications. The operational semantics of CHR is given by a state transition system where states are goals. 
It relies on an equivalence between states that abstracts from the representation of built-ins \cite{raiser_betz_fru_equivalence_revisited_chr09,betz2014unified}.
Basically, two states are equivalent if their built-ins are logically equivalent (imply each other) and 
their user-defined constraints form equivalent multisets taking into account the built-ins.
For example, 
$$X{\leq}Y \land Y{\leq}X \land c(X,Y) \ \equiv \ X{=}Y \land c(X,X) \not\equiv X{=}Y \land c(X,X) \land c(X,X).$$

Let $\CT$ be a (decidable) constraint theory for the built-ins
including equality $=$ over terms with arithmetic expressions.
This means that arithmetic functions are interpreted and 
all other functions are not interpreted, ie. treated syntactically.
A {\em copy (fresh variant, renaming)} of a rule is obtained by uniformly replacing its variables by new variables.
We then say that the variables have been {\em renamed apart}.
\begin{definition}[State Equivalence
\cite{raiser_betz_fru_equivalence_revisited_chr09}]\label{def:stateeq}
{
{\em States} are goals.
Let $C_i$ be the built-ins, let $B_i$ denote user-defined constraints, and let ${\cal V}$ be a set of variables.
Variables of a state that do not occur in ${\cal V}$ are called {\em local variables of the state}.
Two states $S_1 = (C_1 \land B_1)$ and $S_2 = (C_2 \land B_2)$
with local variables $\bar x$ and $\bar y$, respectively, that have been renamed apart,
are {\em equivalent}, 
written $S_1 \equiv_{\cal V} S_2$, if and only if
$$	\CT 
\models 
        \forall (C_1 \rightarrow \exists \bar y ((B_1 = B_2) \land C_2))
	\land 
         \forall (C_2 \rightarrow \exists \bar x ((B_1 = B_2) \land C_1))%
\footnote{This definition implies 
$\CT \models 
\forall (\exists \bar x (B_1 \land C_1) \leftrightarrow \exists \bar y (B_2 \land C_2)).$}.
$$
} 
\end{definition}
$B_1$ and $B_2$ are the multisets of user-defined constraints. They must be pairwise equivalent as enforced by $B_1 = B_2$.
Note that in $B_1$ and $B_2$ we can freely replace a term $t_1$ by another term $t_2$ if the built-ins imply $t_1 = t_2$.
Also, local variables occurring only in the built-ins can be removed (and introduced) if logical equivalence is maintained.
Finally, all states with inconsistent built-ins are equivalent.
These properties have been proven in \cite{raiser_betz_fru_equivalence_revisited_chr09}.
An example illustrates some properties of state equivalence:
$$X{=}Y \land c(X,Y) \equiv_{\{X\}} c(X,X) 
\mbox{ but } 
X{=}Y \land c(X,Y) \not\equiv_{\{X,Y\}} c(X,X).$$

Using this state equivalence, the abstract CHR semantics is defined by a single transition
(computation step) between states. It defines the application of a rule. 
If the source state can be made equivalent to a state that contains the head and the guard of a copy of a rule, then we can apply the rule by replacing the head by the body in the state. Any state that is equivalent to this target state is also in the transition relation.
\begin{definition}[Transition and Computation]\label{transition}
{
A CHR {\em transition (computation step)}
$S \mapsto_r T$ is defined as follows, where $S$ is called {\em source state} and $T$ is called {\em target state}:
\begin{center}
$\underline{S \equiv_{\cal V} (H \land C \land G)  \not\equiv \false \  \ (r : H \Leftrightarrow C \, |\, B) \ \ \ (C \land B \land G) \equiv_{\cal V} T}$\\
$S \mapsto_r T$
\end{center}
where the rule $(r : H \Leftrightarrow C \, |\, B)$ is a copy 
of a rule from a given program $\cal{P}$
such that its local variables do not occur in $G$. %
The goal $G$ is called {\em context} of the rule application. It remains unchanged. 
It may be empty.

A {\em computation (derivation)} of a {\em query (given goal, call)} $S$ with variables ${\cal V}$ in a program $\mathcal{P}$
is a connected sequence
$S_i \mapsto_{r_i} S_{i+1}$ beginning with
the query $S$ as {\em initial state} $S_0$
and either ending in a {\em final state (answer, result)} $S_n$ or 
otherwise {\em not terminating (diverging)}.
The relation $\mapsto^*$ denotes the reflexive and transitive closure of $\mapsto$.
} 
\end{definition}
For convenience, we may drop the reference to 
the rules from the transitions.
We may also drop ${\cal V}$ from the equivalence. 
Note that CHR is a committed-choice language, unlike Prolog there is no backtracking or undoing of rule applications.

\begin{example}[Summation, Contd.]
Recall the rules for summation with $sum/2$:
\begin{gather*}
  b: sum(N,S) \simp N=1 \, |\, S=1\\
  r: sum(N,S) \simp N>1 \, |\, sum(N{-}1,S1), S \aeq N{+}S1
\end{gather*}
Then a computation for the query $sum(3,R)$ proceeds as follows.
\begin{gather*}
sum(3,R)  \, \equiv_{\{R\}}\\
  \, sum(N',S'), N'>1, N'{=}3, S'{=}R \mapsto_{r}\\
N'>1, sum(N'{-}1,S1'), S' \aeq N'{+}S1', N'=3, S'=R  \, \equiv_{\{R\}}\\
  \, sum(3{-}1,S1'), R \aeq 3{+}S1' \mapsto_{r}\\ 
sum(2{-}1,S1'), S1 \aeq 2{+}S1', R \aeq 3{+}S1 \mapsto_{b}\\
S1'{=}1, S1 \aeq 2{+}S1', R \aeq 3{+}S1  \, \equiv_{\{R\}}\\
  \, R{=}6
\end{gather*}
\end{example}

\section{Runtime Repeated Recursion Unfolding}\label{rrrun}

We recall a definition of rule unfolding in CHR.
Next we define simplification inside rule bodies. Then we have all the ingredients necessary to introduce runtime repeated recursion unfolding
and show its correctness. 
We also prove some useful lemmas.
We also show that there is a straightforward optimal rule application strategy and prove it sound and complete.

We will need the standard notions of substitutions, matching and instances.
A {\em substitution} is a mapping function from variables to terms 
$\theta: {\cal V} \rightarrow {\cal T}$, written in postfix notation,
such that domain of $\theta$,
the set $dom(\theta) = \{X \mid X\theta \neq X\}$, is finite.
When a substitution is applied to a goal, 
it is applied to all variables in the goal.
If $A=B\theta$, where $B$ is a goal, 
we say that $A$ is an {\em instance} of $B$,
$A$ {\em matches} $B$, and that $B$ is {\em instantiated}.

\subsection{Rule Unfolding}

For unfolding of rules in CHR, we follow the definition and proofs of \cite{gabbrielli2015unfolding}.
In this paper we rewrite their definition of unfolding in terms of generalized simplification rules. This simplifies the definition and is sufficient for our purposes. 

To define unfolding, we need the following notation.
For a goal $A$,
let $\mathit{vars}(A)$ denote the set of variables in $A$.
Set difference $C_1 = C_2 \setminus C_3$ for conjunctions of built-ins is defined
as $C_1 = \{c \in C_2 \mid \mathcal{CT} \not \models C_3 \rightarrow c\}$.
In words, to obtain $C_1$, remove from $C_2$ the built-in constraints that $C_3$ implies.
\vspace{-1ex}
\begin{definition}[Unfolding (based on Def. 8 \cite{gabbrielli2015unfolding})]\label{def:unf}
{
Let ${\cal P}$ be a CHR program and let $r, v \in {\cal P}$ be two
rules whose variables have been renamed apart
\vspace{-1ex}
$$
\begin{array}{lcl}
r: H & \Leftrightarrow &  C\,|\, D \land B \land G\\
v: H' & \Leftrightarrow & C'\,|\, B',
\end{array}
$$
where $D$ is the conjunction of the built-ins in the body of $r$. 
Then we define the \emph{unfolding of rule $r$ with rule $v$}
\vspace{-1ex}
$$\mathit{unfold}(r,v) = r'$$ 
as follows.
Let  $\theta$ be a substitution such that 
$dom(\theta) \subseteq \mathit{vars}(H')$.
Let $C''\theta = C'\theta \setminus (C \wedge D)$.
If
$\mathcal{CT} \models \exists (C \wedge D)
\land
\forall ((C \wedge D) \rightarrow G{=}H'\theta)$,
$\mathit{vars}(C''\theta) \cap \mathit{vars}(H'\theta) \subseteq \mathit{vars}(H)$ 
and $\mathcal{CT} \models \exists (C \wedge C''\theta)$,
then the unfolded rule $r'$ is
$$r': H \Leftrightarrow C\land C''\theta \, |\, D \land B \land G{=}H' \land B'.$$ 
} 
\end{definition}
If a goal $G$ in the body of rule $r$ matches the head $H'$ of a rule $v$,
unfolding replaces $G$ by the body of rule $v$ together with $G{=}H'$ to obtain a new rule $r'$. We also add to its guard $C$ an instance of a part of the guard of rule $v$. 
This part $C''$ contains the non-redundant built-ins of $C'$ (they are not implied by the built-ins in the rule $r$). 

Note that for a correct unfolding according to the above definition,
three conditions have to be met.
The chosen substitution must make $H'$ equivalent to the matching $G$
in the context of the built-ins of rule $r$ that must be satisfiable.
Under this substitution, the common variables of $H'$ and $C''$ must already occur in $H$,
and finally the guard of the unfolded rule must be satisfiable.
If these conditions are violated, unfolding cannot take place and no unfolded rule is produced.

Correctness of unfolding means that the unfolded rule does not lead to new states when it is applied, it is redundant. 
\begin{lemma}[Correctness of Unfolding]\label{Corrunf}
Given a CHR program with rules $r$ and $v$ and their unfolding resulting in rule $r' = \mathit{unfold}(r,v)$
and a computation with a transition that applies the unfolded rule
$G \mapsto_{r'} G'$.
Then there exists a computation where we replace the transition by a sequence of transitions without the unfolded rule $G \mapsto^{*} G'$ and leave all other states and transitions unchanged.
\end{lemma}
\begin{shortproof}
Correctness of unfolding is proven in 
{\em Corollary $1$} \cite{gabbrielli2015unfolding}. 
\end{shortproof}
In that sense, a correctly unfolded rule is always redundant (but of course its application is expected to improve efficiency).
\begin{lemma}[Redundancy of Unfolded Rules]\label{Redundancy}
Given the rules $r$ and $v$ and their unfolding resulting in rule 
$r' = \mathit{unfold}(r,v)$ 
and any goal $G$ with a transition with the unfolded rule
$$G \mapsto_{r'} G'',$$
then there exist transitions with the original rules
either of the form
$$G \mapsto_r G' \mapsto_v G'' \mbox{ or } G \mapsto_r G'' \equiv \false.$$
\end{lemma}
\begin{shortproof}
The lemma corresponds to {\em Proposition $6$} in the appendix of
\cite{gabbrielli2015unfolding}, where the proof can be found.
\end{shortproof}

\begin{example}[Summation, contd.]\label{sumunf}{
We unfold the recursive rule for summation with (a copy of) itself:
\begin{gather*}
r: sum(N,S) \simp N>1 \, |\, S \aeq N{+}S1, sum(N{-}1,S1)\\
v: sum(N',S') \simp N'>1 \, |\, S' \aeq N'{+}S1', sum(N'{-}1,S1')
\end{gather*}
Then the unfolded rule is
\begin{gather*}
r_1 : sum(N,S) \simp N>1, N{-}1>1 \, |\, S \aeq N{+}S1, sum(N{-}1,S1){=}sum(N',S'),\\ 
S' \aeq N'{+}S1', sum(N'{-}1,S1')
\end{gather*}
Unfolding is possible since its three conditions are met.
First, $sum(N{-}1,S1)$ is an instance of $sum(N',S')$, 
more precisely
$$(N>1, S \aeq N+S1) \rightarrow sum(N{-}1,S1)=sum(N',S')\theta,$$
where the substitution $\theta$ maps $N'$ to $N{-}1$ and $S'$ to $S1$.
Second, 
$$\mathit{vars}(N{-}1>1) \cap  \mathit{vars}(sum(N{-}1,S1)) \subseteq \mathit{vars}(sum(N,S))$$ 
holds since $\{N\}  \cap  \{N,S1\}\subseteq \{N,S\}.$
Third, the new guard $N>1, N{-}1>1$ is satisfiable.
} 
\end{example}
Obviously we can simplify the built-ins of the guard and the body of this rule, and we will define this kind of simplification next.

\subsection{Rule Simplification}\label{rulesimp}

Speedup crucially depends on the amount of simplification that is possible in the unfolded rules. 
We want to replace built-ins by semantically equivalent ones that can be executed more efficiently.
We define a suitable notion of rule simplification and prove it correct.
In this subsection, we basically follow \cite{reclopstr}.
\begin{definition}[Rule Simplification]\label{def:simp}
Given a rule $r$ of the form
$$r: H \Leftrightarrow C \, |\, D \land B,$$
where $D$ are the built-ins and $B$ are the user-defined constraints in the body of the rule.
We define
\begin{gather*}
\mathit{simplify}(r) = 
(H' \Leftrightarrow C' \, |\, D' {\setminus} C' \land B') \mbox{ such that}\\
(H \land C) \equiv_{{\cal V}} (H' \land C')
 \mbox{ and } (C \land D \land B) \equiv_{{\cal V}} (D' \land B'),
\end{gather*}
where $C'$ and $D'$ are the built-ins and $H'$ and $B'$ are the user-defined constraints,
where ${\cal V} = \mathit{vars}(H) \cup \mathit{vars}(H')$,
and $\mathit{simplify}(r)$ is simpler than $r$ according to some strict partial order.
\end{definition}
In the given rule, we replace head and guard, and the body, respectively, by simpler yet state equivalent goals.
We may remove redundant constraints, we may replace constraints by more efficient ones.
The choice of ${\cal V}$ allows us to remove local variables if possible, i.e those that occur only in the guard or body of the rule.
We temporarily add the guard $C$ when we simplify the body for correctness and to improve the simplification. What is simpler depends on the particular built-in constraints used.

For correctness we have to show that the same transitions $S \mapsto T$ are possible with rule $r$ and rule $\mathit{simplify}(r)$.
\begin{theorem}[Correctness of Rule Simplification]\label{corsimp}
(Theorem 1 of \cite{reclopstr})
Let $r = (H \Leftrightarrow C \, |\, D \land B)$ be a rule and
let $s = (H' \Leftrightarrow C' \, |\, D' {\setminus} C' \land B')$ be the simplified rule $s = \mathit{simplify}(r)$.
For any state $S$ and variables ${\cal V}$, $S \mapsto_r T$ iff $S \mapsto_s T$.
\end{theorem}
\begin{proof}
According to the definition of a CHR transition (Def. \ref{transition})
and of rule simplification (Def. \ref{def:simp}), 
we know that
\begin{center}
$S \mapsto_r T \mbox{ iff }
S \equiv_{\cal V} (H \land C \land G)  \not\equiv \false \mbox{ and } (C \land D \land B \land G) \equiv_{\cal V} T$\\
$S \mapsto_s T \mbox{ iff }
S \equiv_{\cal V} (H' \land C' \land G')  \not\equiv \false \mbox{ and } (C' \land D' {\setminus} C' \land B' \land G') \equiv_{\cal V} T$\\
$(H \land C) \equiv_{\cal V'} (H' \land C')$\\
$(C \land D \land B) \equiv_{\cal V'} (D' \land B'),$
\end{center}
where ${\cal V'} = \mathit{vars}(H) \cup \mathit{vars}(H')$.

It suffices to show that $S \mapsto_r T$ implies $S \mapsto_s T$, since the implication in the other direction is symmetric and can be shown in the same way.

Hence we have to show that there exists a goal $G'$ such that
\begin{center}
$S \equiv (H \land C \land G) \equiv_{\cal V} (H' \land C' \land G') \mbox{ if }
(H \land C) \equiv_{\cal V'} (H' \land C')$ and\\
$T \equiv (C \land D \land B \land G) \equiv_{\cal V} (C' \land D' {\setminus} C' \land B' \land G')  \mbox{ if } (C \land D \land B) \equiv_{\cal V'} (D' \land B').$
\end{center}
We choose $G' = C \land G$.
Note that $(C' \land D' {\setminus} C')$ is just $(C' \land D')$.
The main part of the proof reasons on the first-order logic formulas resulting from applying the definition of state equivalence (Def. \ref{def:stateeq}) to the above equivalences. The full proof can be found in appendix A of the full version of \cite{reclopstr}.
%
%
\end{proof}

We conclude this subsection by simplification of the unfolded rule of our running example.
\begin{example}[Summation, contd.]{
Recall the unfolded rule
\begin{gather*}
sum(N,S) \simp N{>}1, N{-}1{>}1 \, |\, S{\aeq}N{+}S1, sum(N{-}1,S1){=}sum(N',S'),\\ 
S'{\aeq}N'{+}S1', sum(N'{-}1,S1').
\end{gather*}
For the head and guard we have that
$$sum(N,S), N{>}1, N{-}1{>}1 \equiv_{\{S,N\}} sum(N,S), N{>}2.$$
For the body we have that
\begin{gather*}
N{>}1, N{-}1{>}1, S{\aeq}N{+}S1, sum(N{-}1,S1){=}sum(N',S'), S'{\aeq}N'{+}S1', sum(N'{-}1,S1') 
\equiv_{\{S,N\}}\\ 
N{>}2, S{\aeq}2{*}N{-}1{+}S1', sum(N{-}2,S1').
\end{gather*}
Thus the unfolded rule can be simplified into the rule
$$sum(N,S) \simp N{>}2 \, |\, S{\aeq}2{*}N{-}1{+}S1', sum(N{-}2,S1').$$
} 
\end{example}

\subsection{Runtime Repeated Recursion Unfolding}

We can now define our novel program optimization strategy of runtime repeated recursion unfolding based on rule unfolding and rule simplification. We prove it correct by showing the redundancy of unfolded recursive rules and their termination.
On the way, we will also prove lemmas
about the number of recursive steps covered and the number of rules generated.

In our method, we start from a call (query) for a CHR constraint defined by a recursive rule.
We unfold the recursive rule with itself and simplify it. 
Then we unfold the resulting rule.
We repeat this process as long as the resulting rules are applicable to the query.
In this paper, we assume a single linear direct recursive rule.

\begin{definition}[Runtime Repeated Recursion Unfolding]\label{def:rru}
{
Let $r$ be a recursive rule and $G$ be a goal. 
Let $$\mathit{unfold}(r) = \mathit{unfold}(r,r).$$
The {\em runtime repeated recursion unfolding} of a recursive rule $r$ with goal $G$ and with rule simplification
is a maximal sequence of rules $r_0, r_1, \ldots$ where
\begin{gather*}
r_0 = r\\ 
r_{i+1} = \mathit{simplify}(\mathit{unfold}(r_i))\mbox{ if } G \mapsto_{r_{i+1}} G', \ (i\geq 0)%
\end{gather*}
} 
\end{definition}
The definition describes the repetition of the following step to produce the desired sequence of more and more unfolded rules:
We unfold and simplify the current unfolded rule $r_i$.
If the unfolding is possible and if the resulting rule 
$r_{i+1}$ is applicable to the query $G$ (as expressed by $G \mapsto_{r_{i+1}} G'$), we add the new rule to the sequence and continue with it.

\vspace{-1ex}
\begin{example}[Summation, contd.]{
Consider a query $sum(10,R)$.
Recall the unfolded simplified rule
$$r_1 = sum(N,S) \simp N{>}2 \, |\, S \aeq 2{*}N{-}1{+}S1, sum(N{-}2,S1).$$
Since $sum(10,R) \mapsto_{r_{1}} 10=N, N{>}2,\ldots$,
we repeat the unfolding:
\begin{gather*}
\mathit{unfold}(r_1) = sum(N,S) \simp N{>}2, N{-}2{>}2 \, |\, S \aeq 2{*}N{-}1{+}S1,\\
sum(N{-}2,S1){=}sum(N',S'), 
S' \aeq 2{*}N'{-}1{+}S1', sum(N'{-}2,S1').
\end{gather*}
The unfolded rule can be simplified into the rule
$$\mathit{simplify}(\mathit{unfold}(r_1)) = r_2 = sum(N,S) \simp N{>}4 \, |\, S \aeq 4{*}N{-}6{+}S1', sum(N{-}4,S1').$$
The rule $r_2$ is applicable to the goal.
Further recursion unfolding results in rules with guards $N>8$ and then $N>16$.
To the latter rule, the goal $sum(10,R)$ is not applicable anymore. 
Hence runtime repeated recursion unfolding stops.
The rules for the goal $sum(10,R)$ are therefore (more unfolded rules come first):
\begin{gather*}
r_3 = sum(N,S) \simp N>8 \, |\, S \aeq 8*N{-}28+S1, sum(N{-}8,S1)\\
r_2 = sum(N,S) \simp N>4 \, |\, S \aeq 4*N{-}6+S1, sum(N{-}4,S1)\\
r_1 = sum(N,S) \simp N>2 \, |\, S \aeq 2*N{-}1+S1, sum(N{-}2,S1)\\
r = r_0 = sum(N,S) \simp N>1 \, |\, S \aeq N+S1, sum(N{-}1,S1)\\
b = sum(N,S) \simp N=1 \, |\, S=1.
\end{gather*}
Note that to the goal $sum(10,R)$ we can apply any of the recursive rules. The most efficient way is to start with the first, most unfolded rule. It covers more recursive steps of the original recursive rule than any other rule.
We will formalize such optimal rule applications in the next section.
} 
\end{example}

We now prove some useful properties of runtime repeated recursion unfolding.
Unfolded recursive rules are redundant. As is the case for any unfolded rule, their computations can also be performed with the original rule.
\begin{lemma}[Redundancy of Unfolded Recursive Rules]\label{RecRedundancy}
Assume a {runtime repeated recursion unfolding} of a recursive rule $r$ with goal $G$. 
It results in a sequence of rules $r_0, \ldots, r_i, \ldots$ where $r=r_0,\ i\geq 0$.

Then for any goal $B$ with a transition
$B \mapsto_{r_{i+1}} B''$
there exist transitions
either of the form
$$B \mapsto_{r_{i}} B' \mapsto_{r_{i}} B'' \ \mbox{  or  } \ B \mapsto_{r_{i}} B'' \equiv \false.$$
\end{lemma}
\vspace{-3ex}
\begin{shortproof}
This claim follows immediately from {\em (Lemma \ref{Redundancy})}.
\end{shortproof}

One computation step (transition) with an unfolded rule corresponds to two computation steps with the rule that was unfolded (if no inconsistency is involved). 
So each unfolded rule doubles the number of recursive steps of the original rule that it covers.
\begin{lemma}[Recursive Steps Covered by Unfolded Recursive Rules]\label{Steps}
Assume {runtime repeated recursion unfolding} of a recursive rule $r$ with goal $G$. 
It results in a sequence of rules $r_0, \ldots, r_i, \ldots$ where $r=r_0, i\geq 0$.
If 
$$G \mapsto_{r_{i}} G' \mbox{ with } G' \not\equiv \false,$$
then there exists a sequence of $2^i$ transitions with rule $r$
$$G \mapsto_{r} G_1 \ldots \mapsto_{r} G_{2^i} \mbox{ with } G' \equiv G_{2^i}.$$
\end{lemma}
\begin{proof}
By correctness of rule simplification {\em (Theorem \ref{corsimp})}, 
rule $r$ and its simplification 
$\mathit{simplify}(r)$ admit equivalent transitions. We can therefore ignore the application of rule simplification (cf. {\em Definition~\ref{def:rru}}) in this proof.

We will use
induction over the rule index $j \ (i \geq j \geq 0)$, going from the largest unfolded rule $r_i$ to the original rule $r_0$.
We actually prove a more general result: 
that with rule $r_j$ we need $2^{i-j}$ transitions.
We first consider the base case.
Our claim holds trivially for $j=i$ resulting in $2^0$, i.e. one transition with rule $r_i$.

For the induction argument, we assume for rule $r_{j+1}$ we need $2^{i-(j+1)}$ transitions. Then for rule $r_j$ we claim to need twice as many, $2^{i-j}$ transitions. This can be shown 
by replacing each transition
$B \mapsto_{r_{j+1}} B''$
by the two transitions
$B \mapsto_{r_{j}} B' \mapsto_{r_{j}} B''$
according to {\em (Lemma \ref{RecRedundancy})}.

The lemma also admits another possible replacement 
$B \mapsto_{r_{i}} B'' \equiv \false$.
But all states in any computation starting with $G$ and ending in $G' \equiv G_{2^i}$ are different from $\false$ because $G' \not\equiv \false$ and no transition is possible from a state $\false$.
So the replacement involving $\false$ is not possible.

Thus for $j=0$, i.e. rule $r_0=r$, we need $2^{i}$ transitions for one transition with rule $r_i$.
\end{proof}
Hence rule $r_i$ covers $2^i$ recursive steps of the original recursive rule $r$ with goal $G$
if the computation does not end in a state $\false$.

For the upcoming lemmas, we define when a goal $G$ takes $n$ {\em recursive steps} with the original recursion.
\begin{definition}[Recursion Depth of a Goal]
Given a goal $G$ with a recursive rule $r$.
Let $n$ be the maximum number of transitions starting from the query $G$ that only involve applications of the given recursive rule $r$
$$G \mapsto_{r} G_1 \ldots \mapsto_{r} G_{n} \mbox{ and there is no transition with $r$ from $G_{n}$.}$$ 
If the computation is finite and terminates, %
then we call $n$ {\em the recursion depth of goal $G$ with rule $r$}.
\end{definition}

We can unfold rules as long as the number of recursive steps they cover does not exceed $n$. This gives us a limit on the number of rules that we can generate.

\begin{lemma}[Number of Unfolded Recursive Rules]\label{Number}
Given a goal $G$ with a recursive rule $r$ that has recursion depth $n$ and ends in a state $G_{n} \not\equiv \false$.
Then repeated recursion unfolding will generate $k$ rules such that $2^k \leq n$. Hence, $k \leq \lfloor \log_2(n) \rfloor$.
\end{lemma}
\begin{proof}
By contradiction:
Assume repeated recursion unfolding generates a rule $r_k$ such that $2^k > n$.
According to {\em Lemma \ref{Steps}} rule $r_k$ allows for a transition with $G$ that is equivalent to $2^k$ transitions with the original recursive rule $r$. But the maximum number of transitions possible with $r$ is just $n$.
\end{proof}
Note that fewer rules than $\lfloor \log_2(n) \rfloor$ may be generated because (further) unfolding is not possible if its three conditions are not met.

\begin{lemma}[Termination of Runtime Repeated Recursion Unfolding]\label{Termination}
Given a goal $G$ with a recursive rule $r$ that has recursion depth $n$ and ends in a state $G_{n} \not\equiv \false$.
Then the runtime repeated recursion unfolding of $r$ with $G$ terminates.
\end{lemma}
\begin{shortproof}
Direct consequence of {\em Lemma \ref{Number}}.
\end{shortproof}
So we can ensure that runtime repeated recursion unfolding terminates with a goal if the original recursive rule terminates with that goal.

We give two simple examples for nontermination.
\begin{example}[Nontermination]{
The goal $p(0)$ does not terminate with the recursive rule:
\begin{gather*}
r: p(N) \simp N{\neq}1 \, |\, p(N{-}1).
\end{gather*}
Runtime repeated recursion unfolding with goal $p(0)$ results in the rule
\begin{gather*}
r_1: p(N) \simp N{\neq}1,N{\neq}2 \, |\, p(N{-}2).
\end{gather*}
Since the rule is applicable to the goal $p(0)$, 
our unfolding can proceed. Each unfolding adds an inequality to the guard, but the guards will always admit $N{=}0$. Therefore, runtime repeated recursion unfolding does not terminate as well.
} 
\end{example}
The next example shows that the condition that the resulting state is not 
$\false$ is necessary. 
We use a variation of the rule above.
\begin{example}[Nontermination with $\false$]{
The goal $p(0)$ terminates in a state $\false$ 
when applying the following recursive rule,
\begin{gather*}
r: p(N) \simp N{\neq}1 \, |\, N{<}0, p(N{-}1),
\end{gather*}
since the body built-in $N{<}0$ is inconsistent with $N{=}0$ from the goal $p(0)$.
The unfolded and simplified rule is
\begin{gather*}
r_1: p(N) \simp N{\neq}1,N{\neq}2 \, |\, N{<}0, p(N{-}2).
\end{gather*}
Again, with the goal $p(0)$, unfolding can proceed forever. 
Runtime repeated recursion unfolding does not terminate even though the computation with the original rule $r$ terminated.
Still, for the goal $p(0)$ any computation with any unfolded rule will lead to $\false$.
} 
\end{example}

Based on the lemmas proven, we can now directly show correctness of our method.
\begin{theorem}[Correctness of Runtime Repeated Recursion Unfolding]\label{rrrucorrect}
Given a goal $G$ with a recursive rule $r$ that has recursion depth $n$ and ends in a state $G_{n} \not\equiv \false$.
Then the runtime repeated recursion unfolding of rule $r$ with goal $G$ terminates
and generates redundant unfolded rules.
\end{theorem}
\vspace{-2ex}
\begin{proof}
The claim is a direct consequence of termination proven in {\em Lemma \ref{Termination}} and the redundancy of unfolded recursive rules proven in {\em Lemma \ref{RecRedundancy}}.
\end{proof}

\subsection{Optimal Rule Applications}

An unfolded rule covers twice as many recursion steps than the given rule.
When we apply a more unfolded rule, we cover more recursive steps with a single rule application.
Based on this observation we introduce a rule application strategy where we try 
to apply more unfolded rules first. Furthermore each unfolded rule is tried only once and is applied at most once.
We prove our optimal rule application strategy sound and complete.

\begin{definition}[Optimal Rule Application Strategy]\label{optrule}
{
Given a recursive rule $r_0$ with a goal $G$ with $k$ additional rules
$r_0, r_1, \ldots, r_{k-1}, r_k$
from runtime repeated recursion unfolding.
Let the notation $G_i \mapsto^{opt}_r G'$ be shorthand for $G_i \mapsto_r G'$ 
if $G' \not\equiv \false$ %
or otherwise $G_i \equiv G'$.
Then the {\em optimal rule application strategy} is as follows:
$$ G \mapsto^{opt}_{r_k} G_k \mapsto^{opt}_{r_{k-1}} G_{k-1} \ldots G_2 \mapsto^{opt}_{r_1} G_1 \mapsto^{opt}_{r_0} G_0.$$ 
}\end{definition}
As a result of this strategy, to the query $G$ we apply the most unfolded rule $r_k$ exactly once\footnote{We know the application is possible since otherwise the unfolding would not have taken place.}.
In the remaining computation, no matter if a rule $r_{i} \ (i < k)$ was applied or not, 
we next try to apply rule $r_{i-1}$ until $i{=}0$.

We first show soundness of this computation strategy. Computations with optimal rule applications correspond to computations with the original rule only.
\begin{theorem}[Soundness of Optimal Rule Applications]\label{coroptrule}
Given a recursive rule $r_0$ with a goal $G$ with $k$ additional rules
from runtime repeated recursion unfolding.

Then for a computation for goal $G$ with optimal rule applications 
there exists a computation for $G$ only using the original recursive rule $r$ that ends in an equivalent state.
\end{theorem}
\begin{proof}
In such a computation, by {\em Lemma \ref{RecRedundancy}}, 
we can replace a transition with rule $r_{i+1} \ (0 \leq i<k)$ by transitions with only rule $r_i$.
Furthermore the resulting states of these computations are equivalent.
Thus we can repeat this process of replacement until all transitions only involve rule $r$ and the computation will end in an equivalent state. 
\end{proof}

We have seen that a transition with an unfolded rule can replace transitions with the original rule. The other direction is not necessarily true. The unfolded rule may not be applicable because the guard of the unfolded rule may come out stricter than necessary. 
For our optimal rule application strategy to be complete, we require that
unfolding generates all rules with the following property:
If a rule can perform two recursive transitions for a goal, then 
its unfolded rule is also applicable to the goal. 
\begin{theorem}[Completeness of Optimal Rule Applications]\label{compoptrule}
Let $r_0$ be a recursive rule with a goal $G$ with recursion depth $n$ 
with $k = \lfloor \log_2(n) \rfloor$ rules
from runtime repeated recursion unfolding,
where for any rule $r_i \ (0 \leq i<k)$ and any goal $B$ with transitions
$B \mapsto_{r_{i}} B' \mapsto_{r_{i}} B''$ 
there exists a transition
$B \mapsto_{r_{i+1}} B''$.

Then for any computation for $G$ with rule $r_0$ with recursion depth $n$
there exists a computation for $G$ with optimal rule applications that ends in an equivalent state.
\end{theorem}
\begin{proof}
We start from a computation only using rule $r_0$.
According to the condition in the claim, 
we can replace the first two transitions with rule $r_0$ by one transition with rule $r_1$ without changing the resulting state. We repeat this for the remaining pairs of subsequent transitions. We get a computation with transitions using rule $r_1$ ending in at most one transition with rule $r_0$.
With rule $r_1$ we start from the first transition again and repeat this process of replacing two transitions by one of rule $r_2$. 
We continue going from rule $r_i$ to rule $r_{i+1}$ until $i+1=k$.
But now we have a computation that applies each rule from $r_k, r_{k-1}, \ldots, r_1, r_0$ at most once and in the given order of the rules. So this computation is one with optimal rule applications.
\end{proof}

\begin{example}[Summation, contd.]{
Recall that the rules for $sum/2$ are:
\begin{gather*}
r_3 = sum(N,S) \simp N>8 \, |\, S \aeq 8*N{-}28+S1, sum(N{-}8,S1)\\
r_2 = sum(N,S) \simp N>4 \, |\, S \aeq 4*N{-}6+S1, sum(N{-}4,S1)\\
r_1 = sum(N,S) \simp N>2 \, |\, S \aeq 2*N{-}1+S1, sum(N{-}2,S1)\\
r = r_0 = sum(N,S) \simp N>1 \, |\, S \aeq N+S1, sum(N{-}1,S1)\\ 
b = sum(N,S) \simp N=1 \, |\, S=1.
\end{gather*}
A computation with optimal rule applications for the goal $sum(10,R)$ is:
\begin{gather*}
sum(10,R) \mapsto_{r_3}\\ 
10=N, N>8, R=S, S \aeq 8*N{-}28+S1, sum(N{-}8,S1) \equiv_{\{R\}}\\
R \aeq 52+S1, sum(2,S1) \mapsto_{r_0}\\
R \aeq 52+S1, 2=N', N'>1, S1=S', S' \aeq N'+S1', sum(N'{-}1,S1') \equiv_{\{R\}}\\
R \aeq 54+S1', sum(1,S1') \mapsto_b\\
R \aeq 55
\end{gather*}
} 
\end{example}

\section{Implementation of Runtime Repeated Recursion Unfolding}\label{impl}

We introduce the implementation of our runtime program transformation.
At compile-time, the rules for the given {\em recursive constraint} are replaced by a call to the unfolder that contains these rules and then to the meta-interpreter that interprets the unfolded rules.
At runtime, the unfolder repeatedly unfolds a recursive rule as long as it is applicable to a given goal using a predefined {\em unfolding scheme} that includes the simplification.
Then the meta-interpreter applies the resulting unfolded rules according to the optimal rule application strategy.
We use an interpreter because we do not want to modify the given program at runtime.

For our implementation, we use CHR embedded in Prolog.
Such sequential CHR
systems execute the constraints in a goal from left to right and apply rules top-down according to
their textual order in the program. 
A user-defined constraint in a goal
can be understood as a procedure call that traverses the rules of the program.
If it and possibly previous constraints from the goal match the head of a rule,
a copy of the rule is instantiated according to the matching.
If the guard check of the rule copy holds, then the rule is applicable.
For application, 
the matched constraints are replaced by the body of the rule copy and execution continues with the calls in the body.
The first applicable rule will
be applied, and this application cannot be undone, it is committed-choice (in
contrast to clause application in Prolog).
This behavior has
been formalized in the so-called {\em refined semantics} which is a
proven concretization of the abstract operational semantics \cite{duck_stuck_garc_holz_refined_op_sem_iclp04}.

According to the CHR semantics, all Prolog predicates are regarded as built-in constraints.
In the following code in concrete syntax, 
{\tt =/2}, {\tt copy\_term/2} and {\tt call/1} are standard built-in predicates of Prolog. 
The syntactic equality {\tt =/2} tries to unify its arguments, i.e. making them syntactically identical by instantiating their variables appropriately.
The built-in {\tt copy\_term/2} produces a copy (variant, renaming) of the given term with new fresh variables. 
The arithmetic equality {\tt is/2} tries to unify its first argument with the result of evaluating the arithmetic expression in its second argument.
The Prolog meta-call {\tt call/1} executes its argument as a goal. It works for both Prolog built-in predicates and CHR constraints.
Our implementation with CHR in SWI Prolog \cite{wielemaker2012swi,swi2005chr} together with the examples and benchmarking code is available online at 
\url{https://exia.informatik.uni-ulm.de/fruehwirth/rrru.pl}.

\subsection{Unfolder Implementation}\label{unfimpl}

The unfolder is implemented as a recursive CHR constraint {\tt unf/3}. 
It repeatedly unfolds and simplifies a recursive rule as long as it is applicable to a goal.
In {\tt unf(G,Rs,URs)}, the first argument {\tt G} is the goal and
{\tt Rs} is a list of rules. {\tt URs} is the resulting list of unfolded rules.
We assume that in the goal {\tt G} the input arguments are given and the output arguments are variables.
Initially, the list {\tt Rs} consists of the recursive rule followed by one or more rules for the base cases of the recursion.
Consider the code below.
The comment in the first line declares the arguments of {\tt unf/3} as either input ({\tt +}) or output ({\tt -}).
A variable that occurs only once in a CHR rule has a name that starts with an underscore character.

\eject

\begin{verbatim}
 % unf(+RecursiveGoal, +RuleList, -UnfoldedRuleList)

unf(G, [R|Rs], URs) <=>             % recursive step
   R = (H <=> Co | _),              % get head H and guard Co of rule R
   copy_term((H <=> Co),(G <=> C)), % copy them, unify head copy with goal G
   call(C)     	                    % call instantiated guard copy C
   |
   simp_unf(R, UR),                 % unfold and simplify rule R into rule UR
   unf(G, [UR,R|Rs], URs).          % add new rule UR and recurse

unf(_G, [_R|Rs], URs) <=> URs=Rs.   % otherwise return rules Rs in URs
\end{verbatim}
In a recursive step of {\tt unf/3},
the first rule element in the list {\tt Rs} is unfolded and added in front of {\tt Rs}.
In the base case of the recursion, the final resulting list of unfolded and given rules is returned in {\tt URs}. 
We rely on the refined CHR semantics and its rule order to ensure that the rule for the base case is only applied if the recursive rule is not applicable.

We explain the recursive rule for {\tt unf/3} in detail now.
We check if the rule {\tt R} in the list is applicable to the {query (call, goal)} {\tt G}.
The guard check is performed by getting (using {\tt =/2}) and copying the relevant parts (head and guard) of rule {\tt R}, unifying
the copied head with the goal (all with {\tt copy\_term/2}) and then executing the instantiated guard copy with {\tt call/1}. The copies will not be needed after that.

If the guard check succeeds, 
we unfold the current rule {\tt R} with itself and and simplify it using {\tt simp\_unf/2} and add the resulting rule {\tt UR} to the rule list in the recursive call of {\tt unf/3}.
Note that we unfold the given general rule, not the instance of the rule stemming from the query.

The Prolog predicate {\tt simp\_unf/2} implements the {\em unfolding scheme}.
Its call {\tt simp\_unf(R,UR)} 
takes the current rule {\tt R${=}r_i$} and computes its simplified unfolding {\tt UR${=}r_{i+1}$}
according to $r_{i+1} = \mathit{simplify}(\mathit{unfold}(r_i))$ in Definition \ref{def:rru}.
For ease of implementing {\tt simp\_unf/2}, we use a {\em rule template $t_r$} which is a suitable generalization of the given recursive rule $r{=}r_0$ and its simplified unfoldings $r_i$. The rules are then instances of the template, i.e. $r_i = t_r\theta_i$. 
The substituted variables $dom(\theta_i)$ in the template represent the {\em parameters} for the instance. The parameters will be bound at runtime.
Therefore the head of the clause for {\tt simp\_unf/2} will be of the form {\tt simp\_unf($t_r$,$t'_r$)}.
In the body of {\tt simp\_unf/2}, the parameters for the unfolded rule will be computed from the parameters of the current rule
where $r_i$ and $r_{i+1}$ will be an instance of $t_r$ and $t'_r$. 

When the guard check has failed, the base case of {\tt unf/3} returns the rules that have been accumulated in the rule list as the result list in the third argument (with the exception of the first rule to which the goal was not applicable).

To simplify the implementation, the body of the rules in the lists
syntactically always consists of three conjuncts of goals:
the constraints before the recursive goal, the recursive goal and the constraints after the recursive goal. 
If there are no such constraints (or no recursive goal in the base case), we use the built-in {\tt true} to denote an empty conjunct.

The following example clarifies the above remarks on the implementation.
\begin{example}[Summation, contd.]\label{sumimpl}{
We show how we implement unfold and simplify with {\tt simp\_unf/2}
for the summation example. 
We abbreviate $sum$ to its first letter {\tt s} to avoid clutter in the code. 
The rule template for $sum$ is
\begin{verbatim}
s(A,C) <=> A>V | B is A-V, s(B,D), C is V*A-W+D   % rule template
\end{verbatim}
where the variables {\tt V} and {\tt W} are parameters that stand for integers.
Its instance for the original recursive rule is
\begin{verbatim}
s(A,C) <=> A>1 | B is A-1, s(B,D), C is 1*A-0+D   % rule instance V=1, W=0
\end{verbatim}

The implementation of the unfolding scheme for summation is accomplished by the following Prolog clause for {\tt simp\_unf/2}.
\begin{verbatim}
simp_unf(
  (s(A,C) <=> A>V | B is A-V, s(B,D), C is V*A-W+D), 
  (s(Al,Cl) <=> Al>Vl | Bl is Al-Vl, s(Bl,Dl), Cl is Vl*Al-Wl+Dl)
  ) :-
    Vl is 2*V, Wl is 2*W+V*V.
\end{verbatim}

For a goal {\tt s(100,S)} the unfolder is called with 
\begin{verbatim}
unf(s(100,S), [
     (s(A,C)<=>A>1| B is A-1, s(B,D), C is 1*A-0+D),  % original recursion
     (s(A,B)<=>A=1| B=1, true, true)                  % base case
     ], URs). 
\end{verbatim}
It will return the following rules in the list {\tt URs}:
\begin{verbatim}
s(A,C) <=> A>64 | B is A-64, s(B, D), C is 64*A-2016+D
s(A,C) <=> A>32 | B is A-32, s(B, D), C is 32*A-496+D
s(A,C) <=> A>16 | B is A-16, s(B, D), C is 16*A-120+D
s(A,C) <=> A>8 | B is A-8, s(B, D), C is 8*A-28+D
s(A,C) <=> A>4 | B is A-4, s(B, D), C is 4*A-6+D
s(A,C) <=> A>2 | B is A-2, s(B, D), C is 2*A-1+D
s(A,C) <=> A>1 | B is A-1, s(B, D), C is 1*A-0+D      % original recursion
s(A,C) <=> A=1 | C=1, true, true                      % base case
\end{verbatim}
} 
\end{example}

\subsection{Meta-Interpreter Implementation}

We implement the optimal rule application strategy with the help of a meta-interpreter for CHR. 
Our meta-interpreter handles the recursive calls, any other goal will be handled by the underlying CHR implementation. 
To a recursive goal, the meta-interpreter tries to apply the unfolded rules produced by the unfolder and applies them at most once.
The meta-interpreter is called with {\tt mip(G,Rs)}, where {\tt G} is the given recursive goal and {\tt Rs} is the list of rules from the unfolder {\tt unf/3}.

\eject

\begin{verbatim}
 % mip(+RecursiveGoal, +RuleList)

mip(true,_Rs) <=> true. % base case, no more recursive goal
mip(G,[R|Rs]) <=>       % current rule is applicable to goal G
   copy_term(R, (G <=> C | B,G1,D)), % copy rule, unify head copy with G
   call(C)              % check guard
   |
   call(B),             % execute constraints before recursive call
   mip(G1,Rs),          % recurse with recursive goal and remaining rules
   call(D).             % execute constraints after recursive call
mip(G,[_R|Rs]) <=>      % current rule is not applicable 
    mip(G,Rs).          % try remaining rules on G
\end{verbatim}
We now discuss the three rules of our meta-interpreter.
\begin{itemize}
\item In the first rule, the base case is reached since the recursive goal has been reduced to {\tt true}.

\item The second meta-interpreter rule 
tries to apply the rule {\tt R} in the rule list to the current goal {\tt G}. It copies the rule, unifies the copied head with the goal and then checks if the guard {\tt C} holds with a meta-call.
If so, the rule is applied.
The conjunct before the recursive goal {\tt B} is directly executed with a meta-call.
Next, the recursive goal {\tt G1} is handled with a recursive call to the meta-interpreter using the remainder of the rule list.
Finally the conjunct after the recursive goal {\tt D} is directly executed with a meta-call.

\item Otherwise the first rule from the rule list was not applicable (according to the refined semantics), and then the last meta-interpreter rule recursively continues with the remaining rules in the list.
\end{itemize}
This ensures that each unfolded rule is tried and applied at most once in accordance with the optimal rule application strategy.

\subsection{Recursive Constraint Implementation}

In order to enable runtime repeated recursion unfolding,
at compile-time, the rules for the given recursive constraint {\tt c/k} are replaced by a call to the unfolder {\tt unf/3} that contains these rules and then to the meta-interpreter {\tt mip/2} that interprets the unfolded rules.
We replace 
according to the rule template named {\tt rec\_unfold} where 
{\tt X1,...,Xk} are different variables and
{\tt OriginalRules} is the list of the given original rules that defined the recursive constraint. 
\begin{verbatim}
 % rule template for a recursive constraint c/k

rec_unfold @ c(X1,...,Xk) <=> 
     unf(c(X1,...,Xk), OriginalRules, UnfoldedRules),
     mip(c(X1,...,Xk), UnfoldedRules).
\end{verbatim}

\begin{example}[Summation, contd.]
For the summation example, the {\tt rec\_unfold} rule instance is as follows:
\begin{verbatim}
rec_unfold @ sum(N,S) <=> 
    unf(s(N,S), [
       (s(A,C) <=> A>1 | B is A-1, s(B, D), C is 1*A-0+D),
       (s(A,C) <=> A=1 | C=1, true, true)
                 ], URs), 
    mip(s(N,S), URs).
\end{verbatim}
\end{example}

\section{Time Complexity of the Implementation}\label{impltime}

For the worst-case time complexity of our implementation of runtime repeated recursion unfolding,
we have to consider the recursion in the original rule, and the recursions in the unfolder as well as meta-interpreter.
We parametrize the time complexity by the number of recursive steps with the original rule. 
From the time complexity of the recursive step we can derive the time complexity of the recursion using recurrence equations.
This gives us a precise measure of the complexity of the recursion.
We assume some familiarity with the T-notation for time complexity (cf. Chapter 2 in \cite{cormen}) as well as stating and solving recurrences (cf. Chapter 4 in \cite{cormen}).

Our time complexity considerations are based on \cite{fruhwirth2002time} and on the following realizable assumptions for the Prolog built-in predicates: 
Matching, unification and copying take constant time for given terms and linear time in the size of the involved terms in general.
A Prolog meta-call has the same time complexity as directly executing its goal argument.

In the following annotated code for the unfolder and meta-interpreter, the comments indicate the time complexity of each non-recursive goal in the bodies of the rules. A comment with symbol {\tt *} in front indicates a non-recursive goal whose execution dominates the complexity of a recursive step.

\subsection{Time Complexity of the Original Rule}

The time complexity of the original recursive rule is straightforward to derive.
\begin{lemma}[Worst-Case Time Complexity of the Original Rule]\label{comorig}
The worst-case time complexity $T_r(n)$ 
of taking $n$ recursive steps with
the given recursive rule $r = (H \Leftrightarrow C \, |\, D \land B \land H')$ can be derived from the the recurrence equation
$$T_r(n) = T_b(n) + T_r(n{-}1),$$ 
where $T_b(n)$ is the time complexity of the $n$-th recursive step $C \land D \land B$ of rule $r$.
\end{lemma}
\begin{proof}
The recurrence follows directly from the structure of the linear direct recursive rule $r$.
\end{proof}

\subsection{Time Complexity of the Unfolder}

For the unfolder we can derive the time complexity of its rules as follows:
\begin{verbatim}
unf(G, [R|Rs], URs) <=>              % head matching constant time
   R = (H <=> Co | _),               % unification constant time
   copy_term((H <=> Co), (G <=> C)), %*time linear in size of head and guard
   call(C)                           %*time of guard call of rule instance
   |
   simp_unf(R, UR),                  %*time for rule template computation
   unf(G, [UR,R|Rs], URs).           % recursive call

unf(_G, [_R|Rs], URs) <=> URs=Rs.    % matching and unification constant time
\end{verbatim}

The complexity of the rule for the base case is constant.
The complexity of a recursive step mainly depends on the time for copying head and guard, for guard checking, and for unfolding and simplification of the current rule.

\begin{lemma}[Worst-Case Time Complexity of the Unfolder]\label{comunf}
Given a {terminating} goal $G$ that has recursion depth $n$ with the given recursive rule $r$.
Then the worst-case time complexity $T_\mathit{unf}(n)$ of the unfolder {\tt unf/3} for goal $G$ with rule $r$  can be derived from the the recurrence equation
$$T_\mathit{unf}(n) = T_c(n) + T_\mathit{unf}(n/2),$$
where 
$$T_c(n) = c + T_\mathit{copy\_term}(n) + T_\mathit{call\_guard}(n) + T_\mathit{simp\_unf}(n)$$
is the time complexity of a recursive step of the unfolder
with the unfolded rule $r_i$ with $i = \lfloor \log_2(n) \rfloor$.
In the summation, the notation $T_p$ denotes the runtime of predicate $p$ in the given code.
\end{lemma}
\begin{proof}
The base case of the unfolder takes constant time and can therefore be ignored.
The recurrence halves $n$. We show that this is correct.
By Lemma \ref{Number} we know that $k$ unfolded rules will be returned by the unfolder such that 
$2^k \leq n$. 
In each recursive step, 
the unfolder doubles the number of recursive steps covered by the currently unfolded rule and the number will not exceed $n$.
Thus the complexity of generating these rules is the sum of $T_c(2^i)$ with $0 \leq i \leq k$.
On the other hand, the recurrence halves $n$ in each recursive step.
This results in the sum of $T_c(n / 2^j)$ with $0 \leq j \leq \log_2(n)$.
But then for each $T_c(2^i)$ we have a corresponding $T_c(n / 2^j)$ with $j=k-i$ such that $2^i \leq n / 2^j$ since $2^k \leq n$.
Therefore the recurrence provides an upper bound on the time complexity of the rules.

Finally, the definition of the complexity for the recursive step $T_c(n)$ can be directly read off the annotated code for the unfolder given above.
The constant $c$ is the time needed for the head matching and the unification in the guard.
\end{proof}
Note that the number of recursive steps of the unfolder (and meta-interpreter) is logarithmic in the number of recursive steps of the original rule. This also reduces the overhead incurred by unfolding and meta-interpretation.

\subsection{Time Complexity of the Meta-Interpreter}

In the following code for the meta-interpreter, again comments indicate the runtime of each goal.
\begin{verbatim}
mip(true,_Rs) <=> true. % head matching and body execution constant time

mip(G,[R|Rs]) <=>       % head matching constant time
   copy_term(R, (G <=> C | B,G1,D)), %*time linear in rule size
   call(C)              %*runtime of guard C of rule instance
   |
   call(B),             %*runtime of goal B of rule instance
   mip(G1,Rs),          % recursive call
   call(D).             %*runtime of goal D of rule instance

mip(G,[_R|Rs]) <=>      % head matching constant time
    mip(G,Rs).          % recursive call
\end{verbatim}
The second rule of the meta-interpreter applies a rule from the list to the current goal.
It dominates the complexity.
Its complexity is determined by the time needed for copying the rule
and for the meta-calls of the guard and of the two body conjuncts of the rule.
The third rule is also recursive in the rule list. The complexity of its recursive step is constant.
The complexity of the rule for the base case is constant.

The resulting recurrence for complexity and its proof are analogous to the one for the unfolder.
\begin{lemma}[Worst-Case Time Complexity of the Meta-Interpreter]\label{comint}
Given a {terminating} goal $G$ that has recursion depth $n$ with the given recursive rule $r$.
Then the worst-case time complexity $T_\mathit{mip}(n)$ of the meta-interpreter {\tt mip/2} for goal $G$ with rule $r$  can be derived from the the recurrence equation
$$T_\mathit{mip}(n) = T_d(n) + T_\mathit{mip}(n/2),$$
where 
$$T_d(n) = c + T_\mathit{copy\_term}(n) + T_\mathit{call\_guard}(n) + T_\mathit{call\_body}(n)$$
is the time complexity of a recursive step of the second rule of the meta-interpreter
with the unfolded rule $r_i$ with $i = \lfloor \log_2(n) \rfloor$.
The runtime $T_\mathit{call\_guard}(n)$ refers to the complexity of the goal {\tt call(C)} for the guard and $T_\mathit{call\_body}(n)$ to the complexity of the body goals {\tt call(B)} and {\tt call(D)}.
\end{lemma}
\begin{proof}
The base case of the meta-interpreter takes constant time and can therefore be ignored.
The recurrence halves $n$. We show that this is correct.
The unfolder returned $k$ unfolded rules with $2^k \leq n$ (cf. Lemma \ref{Number}).
These rules are ordered such that the more unfolded rules come first. 
In each recursive step, the meta-interpreter tries to apply the current unfolded rule once and then proceeds to the next one.
Rule $r_i$ 
covers $2^i$ recursive steps of the original rule $r$.
Thus the complexity of applying these rules is the sum of $T_d(2^i)$ with $0 \leq i \leq k$.

The remainder of this proof is analogous to the one for the unfolder:
Since the recurrence halves $n$ in each recursive step,
it results in the sum of $T_d(n / 2^j)$ with $0 \leq j \leq \log_2(n)$.
But then for each $T_d(2^i)$ we have a corresponding $T_d(n / 2^j)$ with $j=k-i$ such that $2^i \leq n / 2^j$ since $2^k \leq n$.
Therefore the recurrence provides an upper bound on the time complexity of the rules.

Again, the definition of the complexity for the recursive step $T_d(n)$ can be directly read off the annotated code for the meta-interpreter given above.
The constant $c$ is the time needed for the head matching.
\end{proof}
Note that $T_d(n)$ has about the same time complexity as directly executing the rule (but possibly without optimizations), since the overhead of meta-calls is assumed to be constant and only the cost of copying the rule is added.

\subsection{Time Complexity of Runtime Repeated Recursion Unfolding}

We now can establish the worst-case time complexity of the {recursive constraint}
under runtime repeated recursion unfolding.
Recall that the original rules for the recursive constraint are replaced by the following rule that calls the unfolder and then the meta-interpreter.
\begin{verbatim}
rec_unfold @ G <=> unf(G, Rules, UnfoldedRules), mip(G, UnfoldedRules).
\end{verbatim}

\begin{theorem}[Worst-Case Time Complexity of Runtime Repeated Recursion Unfolding]\label{comrrru}
Given runtime repeated recursion unfolding for the rules of a recursive constraint {\tt G} 
and
the time complexities $T_c(n)$ and $T_d(n)$ for a recursive step of the unfolder and the meta-interpreter, respectively.
Then the worst-case time complexity $T_u(n)$ 
of computing the recursion with runtime repeated recursion unfolding
using the instance of rule {\tt rec\_unfold} for {\tt G}  
can be derived from the the recurrence equation
$$T_u(n) = T_c(n) + T_d(n) + T_u(n/2).$$ 
\end{theorem}
\begin{proof}
Clearly $T_u(n) = T_\mathit{unf}(n) + T_\mathit{mip}(n)$ according to rule {\tt rec\_unfold}. 
Recall the recurrence equations for the worst-case time complexity of the unfolder and meta-interpreter:
\begin{gather*}
T_\mathit{unf}(n) = T_c(n) + T_\mathit{unf}(n/2) \mbox{ (cf. Lemma \ref{comunf})},\\
T_\mathit{mip}(n) = T_d(n) + T_\mathit{mip}(n/2) \mbox{ (cf. Lemma \ref{comint})}.
\end{gather*}
Hence $T_u(n) = (T_c(n) + T_\mathit{unf}(n/2)) + (T_d(n) + T_\mathit{mip}(n/2))$. We can replace $T_\mathit{unf}(n/2)) + T_\mathit{mip}(n/2)$ by $T_u(n/2)$.
Thus $T_u(n) = T_c(n) + T_d(n) + T_u(n/2)$.
\end{proof}

We call $T_c(n) + T_d(n)$ the time complexity of the {\em combined recursive step} of the unfolded recursive constraint.

\section{Super-Linear Speedup Theorems}\label{secspeedup}

We first define a class of time complexities. We then give general tight solutions for the recurrences for the recursions in terms of these complexities. 
In our speedup analysis, we then compare the time complexities of the recursive steps in the original recursion and in runtime repeated recursion unfolding. We establish relationships that lead to a super-linear speedup of the recursion.
Based on them, we prove both a sufficient condition for super-linear speedup as well as a sufficient and necessary condition for super-linear speedup.
For a given recursion, then one tries to find an unfolding and simplification with an improved time complexity that satisfies one of the conditions. If it can be found, a super-linear speedup is guaranteed.

In the following, we assume some familiarity with the 
$\Theta$-notation for complexity and its manipulation (cf. Chapter 9.3 in \cite{Knuth} and Chapter 3 in \cite{cormen}). 
The $\Theta$-notation gives us an upper and lower bound on the complexity by ways of a complexity class.

\subsection{Solving the Recurrences for Polylog-Polynomial Time Complexities}

We consider time complexity classes that are expressible by polylog-polynomial functions of the form $n^j \log(n)^k$ in terms of recursion depth $n$ where $j$ and $k$ are non-negative integers\footnote{As we discuss in Section \ref{discuss} this does not preclude exponential complexity in terms of problem size.}.
This includes as special cases 
polynomial complexity ($k = 0$),
linear complexity ($k = 0, j = 1$),
polylogarithmic complexity ($j = 0$),
logarithmic complexity ($k = 1, j = 0$),
and constant complexity ($k = 0, j = 0$).
We can solve the recurrence without a boundary condition, i.e. without an extra equation for the base case of the recursion assuming the base case has constant time complexity (cf. Chapter 4 \cite{cormen}).

\begin{lemma}[Polylog-Polynomial Time Complexities in Runtime Repeated Recursion Unfolding]\label{complexities}
Given a goal $G$ with a recursive rule $r$ that has recursion depth $n$ and the recursive constraint resulting from runtime repeated recursion unfolding.
Consider recursive steps and recursions with polylog-polynomial time complexities
of the form $n^j \log(n)^k$ where $j{\geq}0, k{\geq}0$ are fixed non-negative integers.

Then for the original recursive rule $r$ with time complexity $T_b(n)$ for the recursive step and time complexity $T_r(n)$ for the recursive computation it holds that
$$T_b(n) = \Theta(n^j \log(n)^k) \mbox{ iff } T_r(n) = \Theta(n^{j+1} \log(n)^{k}).$$

Then for runtime repeated recursion unfolding of rule $r$ with time complexity $T_c(n){+}T_d(n)$ for the combined recursive step and time complexity $T_u(n)$ for the recursive computation it holds that
$$T_c(n){+}T_d(n) = \Theta(\log(n)^k) \mbox{ iff } T_u(n) = \Theta(\log(n)^{k+1}) \mbox { and}$$
$$T_c(n){+}T_d(n) = \Theta(n^j \log(n)^k) \mbox{ iff } T_u(n) = \Theta(n^j \log(n)^k)  \mbox { for } j{\geq}1.$$
\end{lemma}
\begin{proof}
There are three claims. 
We first prove their implications to the right.
We start with
$$T_b(n) = \Theta(n^j \log(n)^k) \Rightarrow T_r(n) = \Theta(n^{j+1} \log(n)^{k})$$
based on the recurrence for a recursive computation with the original rule from Lemma \ref{comorig}
$T_r(n) = T_b(n) + T_r(n{-}1)$.
The complexity $\Theta(n^{j+1} \log(n)^{k})$ is clearly an upper bound for $T_r(n)$ since there are $n$ recursive steps and according to the claim $T_r(n) = n \ T_b(n)$.
It remains to prove that the bound is tight, i.e. that it is also a lower bound.
We compute a lower bound as follows:
For the first $n/2$ recursive steps from $n$ to $n/2$ we approximate $T_b(n)$ from below by 
$\Theta((n/2)^j \log_2(n/2)^k)$ and we ignore the contribution of the rest of the recursion. This gives a complexity of $\Theta((n/2) (n/2)^j \log_2(n/2)^k) = \Theta((n/2)^{j+1} (\log_2(n)-1)^k) = \Theta(n^{j+1} \log_2(n)^{k})$ for a fixed $k$. Hence the upper and lower bounds coincide.

The remaining two claims are based on the recurrence for runtime repeated recursion unfolding from Lemma \ref{comrrru}
$T_u(n) = T_c(n) + T_d(n) + T_u(n/2)$.
We next prove
$$T_c(n){+}T_d(n) = \Theta(\log(n)^k) \Rightarrow T_u(n) = \Theta(\log(n)^{k+1})$$.
The complexity
$\Theta(\log(n)^{k+1})$ is an upper bound for $T_u(n)$ since there are $\log_2(n)$ recursive steps and in the claim $T_r(n) = \Theta(\log_2(n) T_b(n))$.
We prove that the bound is also a lower bound.
For the first $\log_2(n)/2$ recursive steps starting from $n$ we approximate 
$T_c(n){+}T_d(n)$ from below by 
$\Theta((\log_2(n)/2)^k)$ and we ignore the contribution of the rest of the recursion. 
This gives a complexity of $\Theta((\log_2(n)/2) (\log_2(n)/2)^k) = \Theta((\log_2(n)/2)^{k+1}) = \Theta(\log_2(n)^{k+1})$ since $k$ is fixed. Hence the upper and lower bounds coincide.

Now for
$T_c(n){+}T_d(n) = \Theta(n^j \log(n)^k) \Rightarrow T_u(n) = \Theta(n^j \log(n)^k)$ for $j{\geq}1$.
Since we have that $T_c(n){+}T_d(n) = T_u(n)$ here, the complexity 
$\Theta(n^j \log(n)^k)$ is clearly a lower bound.
We show the upper bound $T_u(n) \leq c n^j \log_2(n)^k$ with $n \geq 2$ for a suitably chosen constant $c \geq 1$. We will be using induction.
\begin{gather*}
T_u(n) = (T_c(n){+}T_d(n)) + T_u({n/2}) 
= n^j \log_2(n)^k + c\ (n/2)^{j} \log_2(n/2)^k
\leq\\ 
= n^{j} \log_2(n)^k + c\ (n^{j}/2) \log_2(n)^k
= n^{j} \log_2(n)^k (1 + c/2)
\leq c\ n^j \log_2(n)^k \mbox{ if } c \geq 2. 
\end{gather*}

To prove the implications in the other direction for the three claims it suffices to observe that there is a bijective function (identity or increment) between the exponents of the complexity functions of the recursive steps and the recursions. Having proven one direction, it suffices to invert the functions to prove the other direction.
\end{proof}

\begin{table}
\begin{center}{
\begin{tabular}{|l|c|c|}
\cline{1-3}
Time Complexity Class & Rec. Step $T_b(n)$ & Orig. Rec. $T_r(n)$\\
\cline{1-3}
\cline{1-3}
polylog-polynomial $j{\geq}0, k{\geq}0$ & $\Theta(n^j \log(n)^k)$ & $\Theta(n^{j+1} \log(n)^{k})$\\
\cline{1-3}
\end{tabular}
\caption{Polylog-Polynomial Time Complexity Classes of Original Recursion}
\label{comporig}
}\end{center}
\end{table}

\begin{table}
\begin{center}{
\begin{tabular}{|l|c|c|}
\cline{1-3}
Time Complexity Class & Rec. Step $T_c(n){+}T_d(n)$ & Unfold. Rec. $T_u(n)$\\
\cline{1-3}
\cline{1-3}
(poly)logarithmic, constant $j{=}0, k{\geq}0$ & $\Theta(\log(n)^k)$ & $\Theta(\log(n)^{k+1})$\\
\cline{1-3}
linear, (polylog-)polynomial $j{\geq}1, k{\geq}0$ & $\Theta(n^j \log(n)^k)$ & $\Theta(n^j \log(n)^k)$\\
\cline{1-3}
\end{tabular}
\caption{Polylog-Polynomial Time Complexity Classes of Runtime Repeated Recursion Unfolding}
\label{comprrru}
}\end{center}
\end{table}

We summarize the results of the Lemma in Table \ref{comporig} and Table \ref{comprrru}.
Note that for the original recursive rule, the time complexity always increases by a factor of $\Theta(n)$ when going from a recursive step to the complete recursion.
For runtime repeated recursion unfolding,
going from a recursive step to the complete recursion does not increase the 
worst-case time complexity for the classes that are at least linear, and by a factor of $\Theta(\log(n))$ for the polylogarithmic classes.

\subsection{Sufficient Condition for Super-Linear Speedup} 

We have a {\em super-linear speedup} if the time complexity of runtime repeated recursion unfolding is lower than that of the original recursive computation, $\Theta(T_u(n)) \subset \Theta(T_r(n))$.
The time complexities for the recursions depend on the time complexity for the respective recursive steps.
Based on the general solutions for the recurrences we can now derive simple conditions on the complexity of the recursive steps that imply a super-linear speedup for the whole recursion.
The idea then is to find a simplification of the recursive steps in the unfolded recursion that satisfies such a condition. If we succeed, a super-linear speedup is guaranteed.

We first consider a sufficient condition for super-linear speedup, where 
the combined recursive step of the unfolder and of the meta-interpreter has the same time complexity as a recursive step with the original rule.
Since the original recursive constraint takes $n$ steps and the unfolded constraint just about $log_2(n)$ steps, we expect a considerable speedup in that case.
Even though this theorem will be made redundant by our next, more general theorem, it is worth proving, because it sets the stage and applies to practical examples as we will see, where it easily can be checked.

\begin{theorem}[Sufficient Condition for Super-Linear Speedup]\label{speeduptheobest}
Given a goal with $n$ recursive steps and time complexity $T_r(n)$ with the original recursive rule.
Assume runtime repeated recursion unfolding 
with completeness of optimal rule applications (cf. Theorem \ref{compoptrule})
and recursive computations with time complexity $T_u(n)$
of the polylog-polynomial form $n^j \log(n)^k$ where
$j$ and $k$ are fixed non-negative integers.

Then we have a super-linear speedup 
$$\Theta(T_u(n)) \subset \Theta(T_r(n)) \mbox{ if } \Theta(T_c(n) + T_d(n)) = \Theta(T_b(n)),$$
where 
$T_b(n)$ is the time complexity of the $n$-th recursive step with rule $r$.

\myparagraph{Proof}
Because $\log(n){<}n$ where $n>c$ for some fixed constant $c$,
the given condition $\Theta(T_c(n) + T_d(n)) {=} \Theta(T_b(n))$ implies
$\log(n) \Theta(T_c(n) + T_d(n)) \subset n\Theta(T_b(n))$. 
By Lemma \ref{complexities}, we have that
$\Theta(T_r(n)) {=} \Theta(n\ T_b(n))$ and the upper bound
$\Theta(T_u(n)) \subseteq \Theta(\log(n)(T_c(n) + T_d(n)))$.
Therefore $\log(n) \Theta(T_c(n) + T_d(n)) \subset n\Theta(T_b(n))$ implies
$\Theta(T_u(n)) \subset \Theta(T_r(n))$.
$\qed$
\end{theorem}
Table \ref{speedup} gives the complexities when this sufficient condition for super-linear speedup holds.
For constant and polylogarithmic complexity classes, 
a super-linear speedup by the factor $\Theta(n/\log(n))$ is possible,
and for the other polylog-polynomial time complexity classes, 
a super-linear speedup of $\Theta(n)$. 

\begin{table}
\begin{center}{
\begin{tabular}{|l|c|c|c|}
\cline{1-4}
Time Complexity Class & Recursive Steps  & Original & Repeatedly\\
& $T_b(n){=}T_c(n){+}T_d(n)$ & Recursion $T_r(n)$ & Unfolded $T_u(n)$\\
\cline{1-4}
\cline{1-4}
const., (poly)logarithmic $k{\geq}0$ & $\Theta(\log(n)^k)$ & $\Theta(n\log(n)^k)$ & $\Theta(\log(n)^{k+1})$\\
\cline{1-4}
linear, polynomial, & & & \\ 
polylog-polynomial $j{\geq}1, k{\geq}0$ & $\Theta(n^j \log(n)^k)$ & $\Theta(n^{j+1} \log(n)^k)$ & $\Theta(n^j \log(n)^{k})$\\
\cline{1-4}
\end{tabular}
\caption{Time Complexity Classes for Super-linear Speedup with Sufficient Condition}
\label{speedup}
}\end{center}
\end{table}

\subsection{Sufficient and Necessary Condition for Super-Linear Speedup}

Actually, we can already achieve a super-linear speedup if the 
complexity of the combined recursive step of the unfolder and meta-interpreter is lower than that of \emph{all} recursive steps (i.e. the complete recursion) with the original rule. 
We can even show that this conditions is not only sufficient, but also necessary.

\begin{theorem}[Sufficient and Necessary Condition for Super-Linear Speedup]\label{speeduptheo}
Given a goal with $n$ recursive steps with the original recursive rule.
Assume runtime repeated recursion unfolding 
with completeness of optimal rule applications 
and recursions with polylog-polynomial time complexities.

Then we have a super-linear speedup 
$$\Theta(T_u(n)) \subset \Theta(T_r(n)) \mbox{ iff } \Theta(T_c(n) + T_d(n)) \subset \Theta(n \, T_b(n)).$$

\myparagraph{Proof}
By Lemma \ref{complexities}, 
it holds that $\Theta(T_r(n)) {=} \Theta(n\ T_b(n))$.
We know that $\Theta(T_c(n) + T_d(n))$ is of the form $n^{j} \log(n)^{k}$.
By Lemma \ref{complexities}, it holds that 
$\Theta(T_u(n)) {=} \Theta(\log(n)(T_c(n) + T_d(n)))$ if $j = 0$ and
$\Theta(T_u(n)) {=} \Theta((T_c(n) + T_d(n)))$ if $j \geq 1$.
We consider these two cases.

If $j \geq 1$, then it directly follows from the equations in the Lemma that the two statements in our claim $\Theta(T_u(n)) \subset \Theta(T_r(n))$ and $\Theta(T_c(n) + T_d(n)) \subset \Theta(n \, T_b(n))$ are identical.
If $j = 0$, then it holds that
$T_c(n) + T_d(n) = \log(n)^{k}$. 
So using the equations in the Lemma 
our claim becomes
$\Theta(\log(n) \log(n)^{k}) \subset \Theta(n \, T_b(n)) \mbox{ iff } \Theta(\log(n)^{k}) \subset \Theta(n \, T_b(n))$.
Both sides hold since $\Theta(n) \subseteq \Theta(n \, T_b(n))$ and
since the polylogarithmic complexity class is sub-linear, 
i.e. $\Theta(log(n)^i) \subset \Theta(n)$ for any fixed $i$.
$\qed$
\end{theorem}

\begin{table}
\begin{center}{
\begin{tabular}{|l|c|c|}
\cline{1-3}
Time Complexity Class
& Rec.Step $T_b(n)$ & Rec.Step $T_c(n){+}T_d(n)$\\
\cline{1-3}
\cline{1-3}
constant, polynomial, linear $j{\geq}0, k{\geq}0$ & $\Theta(n^j)$ & $\Theta(n^j \log(n)^k)$\\ 
\cline{1-3}
(poly)logarithmic, 
polylog-polynomial $j{\geq}0, k{\geq}1$ & $\Theta(n^j \log(n)^k)$ &  $\Theta(n^{j+1} \log(n)^{k-1})$\\ 
\cline{1-3}
\end{tabular}
\caption{Highest Time Complexity Classes for Super-linear Speedup with Sufficient and Necessary Condition}
\label{speedupnew}
}\end{center}
\end{table}

In Table \ref{speedupnew} we list the highest complexities for a combined recursive step of the unfolded rules that still lead to a super-linear speedup.
If the time complexity of the recursive step of the original recursion 
$T_b(n)$ is of the form $\Theta(n^{j} \log(n)^{k})$ with $k \geq 1$, 
then the highest complexity class for the combined recursive step $T_c(n) + T_d(n)$ is
$\Theta(n^{j+1} \log(n)^{k-1})$.
If
$T_b(n)$ is of the form $\Theta(n^{j})$, 
then the highest complexity classes for the combined recursive step $T_c(n) + T_d(n)$ are of the form
$\Theta(n^{j} \log(n)^{k})$ for any fixed $k$.
This follows from that fact that $\Theta(n^i \log(n)^h) \subset \Theta(n^j \log(n)^k)$ iff 
$[i,h] < [j,k]$
using lexicographic order\footnote{See the more general case in Chapter 9.1, equation (9.6) in \cite{Knuth}.}, because the polylogarithmic class is sub-linear.

\section{Experimental Evaluation: Examples with Benchmarks}\label{secbench}

Our examples will demonstrate that super-linear speedups are indeed possible. 
With sufficient simplification, the time complexity is effectively reduced when applying runtime repeated recursion unfolding. 
In our experiments, we used the CHR library in SWI Prolog Version 6.2.1 running on an
Apple Mac mini 2018 with Intel Core i5 8GB RAM and OS-X 10.14.6.
We use default settings for SWI Prolog (including stack sizes) except for the command line option {\tt -O} which compiles arithmetic expressions.
During multiple runs of the benchmarks we observed a jitter in timings of at most 5\%.
Because the runtime improvement is so dramatic, we can only benchmark small inputs with the original recursion and have to benchmark larger inputs with 
runtime recursion unfolding.

\subsection{Summation Example, Contd.}

We have already unfolded and simplified the recursive rule for summation in Section \ref{rrrun}, Example \ref{sumunf}.
We introduced the implementation in concrete syntax in Section \ref{unfimpl}, Example \ref{sumimpl}.
We now derive estimates for the time complexities for our summation example and then compare them to benchmark results. We will predict and observe a super-linear speedup.

\subsubsection{Complexity}  

Our example deals with arithmetic built-ins.
SWI Prolog uses the GNU multiple precision arithmetic library (GMP), where integer arithmetic is unbounded. 
Comparison and addition have logarithmic worst-case time complexity in the numbers involved. 
Naive multiplication is quadratic in the logarithm.
A variety of multiplication algorithms are used in GMP to get close to linear complexity.
If one multiplies with a power of $2$, the complexity can be reduced to logarithmic. This is the case in our example.
We have confirmed this with some benchmarks in SWI Prolog.

\paragraph{Original Recursion}
The rule for the original recursion for summation is
\begin{verbatim}
s(A,C) <=> A>1 | B is A-1, s(B,D), C is A+D.
\end{verbatim}
All numbers {\tt A,B,C} and {\tt D} are positive integers.
By induction we can show that for a call {\tt s(E,F)} it holds that $({\tt F}/2)^2 \leq {\tt E} \leq {\tt F}^2$.
The most costly arithmetic operation in the recursive step is the addition {\tt C is A+D}. 
The complexity of addition is logarithmic in its operands.
The number {\tt D} is the result of the recursive call {\tt s(B,D)}.
Hence {\tt D} is quadratic in {\tt B} and thus also quadratic in {\tt A}, since {\tt B is A-1}. 
Since {\tt A>1}, the recursion depth $n = {\tt A}-1$.
So the time complexity of a recursive step $T_b(n)$ is the complexity for computing {\tt C is A+D}, which is $\Theta(\log({\tt A}) + \log({\tt D})) = \Theta(\log(n)+\log(n^2))=\Theta(\log(n)+2\log(n))=\Theta(\log(n))$.
Hence the worst-case time complexity for the original recursive computation $T_r(n)$ is $\Theta(n \log(n))$ according to Lemma \ref{complexities}.

\paragraph{Unfolder} 
By Lemma \ref{comunf} the complexity of a recursive step of the unfolder can be derived from
$$T_c(n) = 1 + T_\mathit{copy\_term}(n) + T_\mathit{call\_guard}(n) + T_\mathit{simp\_unf}(n).$$%
Recall the predicate {\tt simp\_unf/2} for summation {\tt s/2} 
\begin{verbatim}
simp_unf(
  (s(A,C) <=> A>V | B is A-V, s(B,D), C is V*A-W+D),    % given rule template
  (s(Al,Cl) <=> Al>Vl | Bl is Al-Vl, s(Bl,Dl), Cl is Vl*Al-Wl+Dl)  % unfolded
  ) :-
    Vl is 2*V, Wl is 2*W+V*V.
\end{verbatim}

Consider the definition of $T_c(n)$.
For the complexity of $T_\mathit{copy\_term}(n)$ and $T_\mathit{call\_guard}(n)$ we observe the following:
Copying head and guard of an unfolded summation rule and checking its guard 
involves the numbers {\tt A} and {\tt V}.
Because of the guard {\tt A>V}, the value of {\tt V} is bounded by {\tt A}, i.e. $n+1$.
The size of an integer is logarithmic in its value.
So $T_\mathit{copy\_term}(n) = \Theta(2\log({\tt A})+\log({\tt V})) = \Theta(\log(n))$
and the comparison in the guard means that $T_\mathit{call\_guard}(n) = \Theta(\log({\tt A})+\log({\tt V})) = \Theta(\log(n))$.

For the complexity $T_\mathit{simp\_unf}(n)$ of {\tt simp\_unf/2}, consider the given rule template.
The input is {\tt A} and the parameters are {\tt V} and {\tt W}.
All variables stand for positive integers. 
For the worst-case time complexity we need bounds on their values.
We already know that {\tt C} and {\tt D} are quadratic in $n$
and that {\tt A} and {\tt V} are bounded by $n+1$.
So the product {\tt V*A} is bounded by $(n+1)^2$.
Due to the computation {\tt C is V*A-W+D}, the parameter {\tt W} is hence bounded by $2\ (n+1)^2$.
The body of the clause for {\tt simp\_unf/2} contains {\tt Vl is 2*V}. Since the first value for {\tt V} in the original recursion in template form is $1$, 
by induction {\tt V} must be a power of $2$.
Overall, the clause body contains an addition
and three multiplications that always involve a power of $2$ ({\tt 2} or {\tt V}). 
So the time complexity of all arithmetic operations is logarithmic in the values involved.
Since all values are positive, bounded by $2 (n+1)^2$ and some values are quadratic in $n$, we arrive at a worst-case time complexity of $T_\mathit{simp\_unf}(n) = \Theta(\log(2 (n+1)^2))=\Theta(\log(n))$. 

Hence the time complexity for a recursive step of the unfolder 
$T_c(n)$ is $\Theta(\log(n))$.

\paragraph{Meta-Interpreter} 
Recall the complexity of a recursive step of the meta-interpreter 
according to Lemma \ref{comint}
$$T_d(n) = 1 + T_\mathit{copy\_term}(n) + T_\mathit{call\_guard}(n) + T_\mathit{call\_body}(n)$$
and recall that the template for unfolded summation rules is
\begin{verbatim}
s(A,C) <=> A>V | B is A-V, s(B,D), C is V*A-W+D.
\end{verbatim}
As with the unfolder, copying an unfolded summation rule can be done in logarithmic time. 
As for executing the guard and the non-recursive goals of the body, we have a comparison, subtractions, an addition and a multiplication in the rule.
The multiplication is with {\tt V}, a power of $2$.
All values of the variables involved are bounded by $2 (n+1)^2$ and some are quadratic in $n$.
So the time complexity for a recursive step of the meta-interpreter 
$T_d(n)$ is $\Theta(\log(n))$ as well.

\paragraph{Complexity of Runtime Repeated Recursion Unfolding}
The overall time complexity for a recursive computation with
runtime repeated recursion unfolding 
$T_u(n)$ is $\Theta(\log(n)^2)$ according to Lemma \ref{complexities}.
The complexity for $T_c(n)$ and $T_d(n)$ is the same as for a recursive step with the original rule $T_b(n)$, namely $\Theta(\log(n))$.
We therefore satisfy the sufficient condition for super-linear speedup according to Theorem \ref{speeduptheobest}.
So with repeated recursion unfolding the worst-case time complexity is reduced from $\Theta(n \log(n))$ to 
$\Theta(\log(n)^2)$.

\subsubsection{Benchmarks}  

Table \ref{benchsum} shows benchmarks results for the summation example.
Times are given in milliseconds.
Experiments that show a runtime of less than 10 milliseconds are the averages of 1000 runs.
The benchmarks confirm the super-linear speedup.
\begin{table}[htb] 
\begin{center}{
\begin{tabular}{l c l}

\begin{tabular}{|l|r|r|r|}
\cline{1-2}
\multicolumn{2}{|c|}{Original Summation} \\
\cline{1-2}
Input $n$  & Time\\
\cline{1-2}
$2^{15}$ & 3 \\
$2^{16}$ & 6 \\
$2^{17}$ & 12 \\
$2^{18}$ & 24 \\
$2^{19}$ & 48 \\
$2^{20}$ & 108 \\ %
$2^{21}$ & 217 \\
$2^{22}$ & Out of stack\\
\cline{1-2}
\end{tabular}

& &

\begin{tabular}{|l|r|r|r|}
\cline{1-4}
\multicolumn{4}{|c|}{Runtime Repeated Recursion Unfolding} \\
\cline{1-4}
Input $n$ & Unfolder & Interpreter & Total Time\\
\cline{1-4}
$2^{25}$ & 0.03 & 0.03 & 0.06 \\
$2^{50}$ & 0.07 & 0.08 & 0.15 \\
$2^{100}$ & 0.18 & 0.18 & 0.36 \\
$2^{200}$ & 0.41 & 0.40 & 0.81 \\
$2^{400}$ & 0.84 & 0.82 & 1.66 \\
$2^{800}$ & 1.80 & 1.72 & 3.52 \\
$2^{1600}$ & 3.72 & 3.65 & 7.37 \\
\cline{1-4}
$2^{25}+1$ & 0.03 & 0.02 & 0.05 \\
$2^{50}+1$ & 0.07 & 0.05 & 0.12 \\
$2^{100}+1$ & 0.18 & 0.10 & 0.28 \\
$2^{200}+1$ & 0.40 & 0.19 & 0.59 \\
$2^{400}+1$ & 0.84 & 0.39 & 1.23 \\
$2^{800}+1$ & 1.76 & 0.80 & 2.56 \\
$2^{1600}+1$ & 3.72 & 1.59 & 5.31 \\
\cline{1-4}
\cline{1-4}
\end{tabular}

\end{tabular}
\caption{Benchmarks for Summation Example (times in milliseconds)}
\label{benchsum}
}\end{center}
\end{table}

\paragraph{Original Recursion}
In each subsequent table entry, we double the input number.
The runtime roughly doubles. 
So the runtime is at least linear.
This is in line with the expected log-linear time complexity $\Theta(n \log(n))$:
since the numbers are so small, addition is fast, almost constant time,
and the runtime is dominated by the linear time overhead of the recursion itself.
For larger numbers, the original recursion runs out of local stack.

\paragraph{Unfolder and Meta-Interpreter}
For runtime repeated recursion unfolding of our summation example, we give the time needed for the unfolding, the time needed for the execution with the meta-interpreter, and the sum of these timings (column 'Total Time').
Because our method has lower time complexity, it was already 5000 times faster than the original recursion for $n{=}2^{21}$.
Hence we start from $2^{25}$ and in each subsequent table entry, we square the input number instead of just doubling it.

The runtimes of the unfolder and meta-interpreter are similar.
For each squaring of the input number, the their runtimes more than double.
The benchmarks results obtained are consistent with the expected complexity of $\Theta(\log(n)^2)$,
e.g. $0.0000002 \log_2(n)^2 + 0.002 \log_2(n)$ for the unfolder. 

\paragraph{Comparing Recursion Depths $2^i$ and $2^i+1$}
In the meta-interpreter, each of the unfolded rules will be tried by matching its head and checking its guard, but not all rules will be necessarily applied.
This may lead to the seemingly counterintuitive behavior that a larger query runs faster than a smaller one.

Out of curiosity, to see how pronounced this phenomena is,
we compare timings for values of $n$ of the form $2^i$ and $2^i+1$. 
Input numbers of the form $2^i+1$ will need exactly one application of the most unfolded rule $r_i$ to reach the base case,
because the following recursive call has the input number 
computed by {\tt B is A-V}
which is $(2^i+1)-2^{i}$, i.e. $1$.
For numbers of the form $2^i$ however, all unfolded rules are applied.
In this case, the most unfolded rule is $r_{i-1}$ (not $r_i$), yielding a recursive call with input $2^i-2^{i-1}$, i.e. $2^{i-1}$.
To this call, the next less unfolded rule $r_{i-2}$ applies and so on.
As a consequence it roughly halves the runtime of the meta-interpreter when going from a query with input number $2^i$ to $2^i+1$. 
The timings for the unfolder stay about the same, because only one more rule is generated for $2^i+1$ (e.g. $n=2^{1600}+1$ generates 1601 rules).

\subsection{List Reversal Example}

The classical program reverses a given list in a naive way.
It takes the first element of the list, reverses its remainder and adds the element to the end of the reversed list. 
The CHR constraint $r(A,B)$ holds if list $B$ is the reversal of list $A$.
\begin{gather*}
r(E, D) \simp E=[C|A] \, |\, r(A, B), a(B, [C], D)\\
r(E, D) \simp E=[] \, |\, D=[].
\end{gather*}
We use Prolog notation for lists.
The term $[C|A]$ stands for a list with first element $C$ and remaining list $A$. 
The built-in $a(X,Y,Z)$ appends (concatenates) two lists $X$ and $Y$ into a third list $Z$. Its runtime is linear in the length (number of elements) of the first list.

\subsubsection{Runtime Repeated Recursion Unfolding}

Our aim is to find the appropriate rule template for the repeated unfolding of the recursive rule with itself.

\paragraph{Unfolding}
We start with unfolding the original recursive rule with a copy of itself:
\begin{gather*}
r(E, D) \simp E=[C|A] \, |\, r(A, B), a(B, [C], D)\\
r(E', D') \simp E'=[C'|A'] \, |\, r(A', B'), a(B', [C'], D').
\end{gather*}
The unfolding substitutes $E'$ by $A$ in the guard and produces
\begin{gather*}
r(E, D){\simp}E=[C|A], A{=}[C'|A'] \, | \,
r(A, B){=}r(E', D'), r(A', B'),\\
 a(B', [C'], D'), a(B, [C], D).
\end{gather*}
This unfolding is correct because its three conditions are satisfied (cf. Def. \ref{def:unf}).
First, $r(A,B)$ is an instance of $r(E',D')$.
The second condition requires
$\mathit{vars}(A{=}[C'|A']) \cap \mathit{vars}(r(A,B)) \subseteq \mathit{vars}(r(E,D))$, 
i.e. $\{A\} \subseteq \mathit{vars}(r(E,D))$. 
This will hold if we consider the guard:
since $r(E, D) \land E{=}[C|A] \equiv r([C|A], D) \land E{=}[C|A]$,
we can replace $E$ by $[C|A]$
and then $\{A\} \subseteq \mathit{vars}(r([C|A],D))$.
Third and finally, the guard $E=[C|A], A=[C'|A']$ is satisfiable.

\paragraph{Simplification}
Now we proceed with rule simplification for unfolded rules (Definition \ref{def:simp}).
We simplify the head and guard by eliminating the local variable $A$.
$$r(E, D), E=[C|A], A{=}[C'|A'] \ \equiv_{\{E,D\}} \ r(E, D), E=[C,C'|A'].$$
For the body we first simplify by eliminating the local variables $A, E'$ and $D'$.
\begin{gather*}
E=[C|A], A{=}[C'|A'], r(A, B){=}r(E', D'), r(A', B'), a(B', [C'], D'), a(B, [C], D)
\equiv_{\{E,D\}}\\
E=[C,C'|A'], r(A', B'), a(B', [C'], B), a(B, [C], D)
\end{gather*}
The insight for improving the time complexity is that
we can merge the two calls to constraint $a/3$ into one if we concatenate their second arguments $[C']$ and $[C]$.
\begin{gather*}
E=[C,C'|A'], r(A', B'), a(B', [C'], B), a(B, [C], D)
\equiv_{\{E,D\}}\\
E=[C,C'|A'], r(A', B'), a(B', [C',C], D).
\end{gather*}

\paragraph{Generalization}
The insight follows from the fact that list concatenation is associative.
Consequently, we can simplify to two append constraints of the form
$a(F, G, D), a(D, A, B)$,
where the list $G$ is sufficiently known,
into $a(F,E,B)$,
where $E$ is the result of computing
$a(G,A,E)$ already during simplification while unfolding.

This kind of simplification gives rise to a rule template of the following form
\begin{gather*}
r(E, D){\simp}E=[C_1,\ldots,C_m|A'] \, | \, r(A', B'), a(B', [C_m,\ldots,C_1], D).
\end{gather*}
We call $[C_1,\ldots,C_m|A']$ an \emph{open list}, 
because it ends in the list variable $A'$.
The open list has size $m$ because can match any list with at least $m$ elements.
The $m$ elements $C_1,\ldots,C_m$ are called \emph{element variables}.
Note that these element variables 
occur in reversed order in the list in the second argument of $a/3$ in the rule body.

\subsubsection{Implementation}

We use concrete syntax now and the built-in {\tt append/3} for $a/3$.

\paragraph{Unfolding with Simplification}
The unfolding scheme for list reversal is implemented with the following Prolog clause for {\tt simp\_unf/2}. 
\begin{verbatim}
simp_unf(
  (r(A,B) <=> A=E | true, r(C,D), append(D,F,B)),    % given rule template
  (r(Al,Bl) <=> Al=El | true, r(Cl,Dl), append(Dl,Fl,Bl))  % unfolded rule
  ) :-
    copy_term((E,C,F),(El,Cc,Fc1)), 
    copy_term((E,C,F),(Ec,Cl,Fc2)), 
    Cc=Ec,
    append(Fc2,Fc1,Fl).  
\end{verbatim}
During unfolding,
in the given rule template,
the variable {\tt E} in the guard will be instantiated with an open list ending in the variable {\tt C}.
The list {\tt F} in {\tt append/3} then consists of the element variables of {\tt E} in reversed order.
In the unfolded rule template, the number of elements in these two lists is doubled and their relationship of reversal is maintained.

The doubling is achieved by copying the guard list {\tt E} together with its end variable {\tt C} and list {\tt F} twice. 
In the first copy, the guard list {\tt El} ends in {\tt Cc}.
In the second copy, list {\tt Ec} ends in {\tt Cl} from the recursive call in the unfolded rule template.
The variable {\tt Cc} is unified with {\tt Ec} from the second copy, thus doubling the number of element variables in {\tt El}.
In this way, we have constructed a guard list {\tt El} with twice as many element variables that ends in {\tt Cl}.

Finally, the lists resulting from copying {\tt F} twice, {\tt Fc1} and {\tt Fc2}, are concatenated in their reversed order by executing {\tt append/3} in the body of the clause during unfolding. The result is the new reversed list {\tt Fl}
in {\tt append/3} in the unfolded rule template.

\paragraph{Recursive Constraint}
For list reversal, the {\tt rec\_unfold} rule is as follows:
\begin{verbatim}
rec_unfold @ rev(I,O) <=> 
    unf(r(I,O), [
       (r(A, E) <=> A=[D|B] | true, r(B, C), append(C, [D], E), 
       (r(A, B) <=> A=[] | B=[], true, true)
                ], URs), 
    mip(r(I,O), URs).
\end{verbatim}
The list in the second argument of {\tt unf/3} contains the original recursive rule and the rule for the base case in appropriate template form.

\paragraph{Unfolded Rules}
The rules that are returned by the unfolder {\tt unf/3} for a query 
with 17 list elements are
\begin{verbatim}
r(A, T) <=> A=[S,R,Q,P,O,N,M,L,K,J,I,H,G,F,E,D|B] | 
        true, r(B, C), append(C, [D,E,F,G,H,I,J,K,L,M,N,O,P,Q,R,S], T).
r(A, L) <=> A=[K,J,I,H,G,F,E,D|B] | 
        true, r(B, C), append(C, [D,E,F,G,H,I,J,K], L).
r(A, H) <=> A=[G,F,E,D|B] | true, r(B, C), append(C, [D,E,F,G], H).
r(A, F) <=> A=[E,D|B] | true, r(B, C), append(C, [D,E], F).
r(A, E) <=> A=[D|B] | true, r(B, C), append(C, [D], E).
r(A, B) <=> A=[] | B=[], true, true.
\end{verbatim}
We see here an increase in \emph{rule size}. 
With each unfolding, the rule size almost doubles because the number of elements in the lists double. 
For a query with $n$ list elements, we unfold $\lfloor \log_2(n) \rfloor$ times.
So the list in the most unfolded rule has not more than $n$ elements.
Therefore the size of all unfolded rules taken together will be proportional to $n$.
Note that this does not increase overall space complexity, since the corresponding input list has $n$ elements.

\subsubsection{Complexity}
We now derive estimates for the time complexities.

\paragraph{Original Recursion}
With the original rule we have $n$ recursive steps for an input list of length $n$. 
The guard of the rule can be checked in constant time.
In the body, {\tt append/3} traverses the list in its first argument.
The time needed is linear in the length of this list, which is $n$. 
So we have 
$T_b(n)=\Theta(n)$ for a recursive step.
This results in the well-known quadratic complexity $\Theta(n^2)$ of naive list reversal. 

\paragraph{Unfolder} 
Recall the template for an unfolded rule of list reversal:
\begin{verbatim}
r(A, B) <=> A=E | true, r(C, D), append(D, F, B). 
\end{verbatim}
The sizes of the lists in the rule are bounded by the length $n$ of the input list.
We now consider a recursive step of the unfolder.
Copying head and guard of an unfolded rule as well as checking its guard has a runtime that is linear in the size of the open list {\tt E}. 
In {\tt simp\_unf} we copy and concatenate the lists {\tt E} and {\tt F}. 
The worst-case time complexity of a recursive step in the unfolder is therefore 
$T_c(n) = \Theta(n).$

\paragraph{Meta-Interpreter}
In the meta-interpreter,
copying an unfolded rule and checking its guard is linear in the size of the open list {\tt E}.
The time for concatenation with {\tt append/3} is linear in the length of 
the list {\tt D}. 
The runtime complexity of a recursive step in the meta-interpreter is therefore 
$T_d(n) = \Theta(n).$

\paragraph{Complexity of Runtime Repeated Recursion Unfolding}
According to Lemma \ref{complexities}, this gives linear complexity $\Theta(n)$ in the input list length $n$ for the unfolder as well as the meta-interpreter and for both of them together.
We therefore satisfy the sufficient condition for super-linear speedup according to Theorem \ref{speeduptheobest}.
With repeated recursion unfolding the complexity is reduced from $\Theta(n^2)$ to $\Theta(n)$. 

\subsubsection{Benchmarks}

Table \ref{benchrev} shows benchmarks results for the list reversal example.
The list sizes $n$ are powers of $2$.
Times are in seconds.
A time measurement of 0.0n means that it was below 0.01 but more than zero.
The experiments confirm the super-linear speedup using runtime repeated recursion unfolding. 

\begin{table}[h] 
\begin{center}{
\begin{tabular}{l c l}

\begin{tabular}{|l|r|r|r|}
\cline{1-2}
\multicolumn{2}{|c|}{Original list reversal} \\
\cline{1-2}
Input $n$  & Time\\
\cline{1-2}
$2^{9}$ & 0.01 \\
$2^{10}$ & 0.04 \\
$2^{11}$ & 0.16 \\
$2^{12}$ & 0.65 \\
$2^{13}$ & 2.88 \\ %
$2^{14}$ & 11.48 \\
$2^{15}$ & 46.36 \\
\cline{1-2}
\end{tabular}

& &

\begin{tabular}{|l|r|r|r|}
\cline{1-4}
\multicolumn{4}{|c|}{Runtime Repeated Recursion Unfolding} \\
\cline{1-4}
Input $n$ & Unfolder & Interpreter & Total Time\\
\cline{1-4}
$2^{13}-1$ & 0.0n & 0.0n & 0.0n \\
$2^{14}-1$ & 0.01 & 0.0n & 0.01 \\
$2^{15}-1$ & 0.01 & 0.01 & 0.02 \\
$2^{16}-1$ & 0.02 & 0.01 & 0.03 \\
$2^{17}-1$ & 0.05 & 0.02 & 0.07 \\
$2^{18}-1$ & 0.09 & 0.04 & 0.13 \\
$2^{19}-1$ & 0.19 & 0.08 & 0.27 \\
\cline{1-4}
$2^{13}$ & 0.01 & 0.0n & 0.01 \\
$2^{14}$ & 0.01 & 0.0n & 0.01 \\
$2^{15}$ & 0.02 & 0.01 & 0.03 \\
$2^{16}$ & 0.05 & 0.01 & 0.06 \\
$2^{17}$ & 0.09 & 0.03 & 0.12 \\
$2^{18}$ & 0.17 & 0.06 & 0.23 \\
$2^{19}$ & 0.36 & 0.14 & 0.50 \\
\cline{1-4}
\cline{1-4}
\end{tabular}

\end{tabular}
\caption{Benchmarks for List Reversal Example}
\label{benchrev}
}\end{center}
\end{table}

\paragraph{Original Recursion}
For the original recursion,
the benchmarks indicate a complexity consistent with the expected $\Theta(n^2)$.
Doubling the list size increases the runtime by a factor of about four.

\paragraph{Unfolder and Meta-Interpreter}
All measured runtimes are consistent with a linear complexity $\Theta(n)$. 
For list size $n = 2^{13}$, runtime repeated recursion unfolding is already 
two orders of magnitude faster than the original recursion.
A list with half a million elements can be reversed in half a second.

\paragraph{Comparing Recursion Depths $2^i-1$ and $2^i$}
To complete the picture,
we give timings for list lengths $n$ of the form $2^i$ and their predecessor numbers $2^i-1$. 
In the meta-interpreter,
the runtime of applying {\em all} unfolded rules (case of $n=2^i-1$) 
is less than of applying just the next larger unfolded rule 
(which has twice the size and complexity) 
(case of $n=2^i$).
The unfolder takes several times longer than the meta-interpreter.
Going from $2^{i-1}$ to $2^i$, the unfolder generates one more rule and 
the time spent doubles.
Overall, going from $2^i-1$ to $2^i$ almost doubles the total runtime.

\subsection{Sorting Example}

The classical insertion sort program sorts the numbers given in a list in ascending order:
\begin{gather*}
s(L, S) \simp L{=}[A|L_1] \, |\, s(L_1, S_1), i(A, S_1, S)\\
s([], S) \simp S{=}[].
\end{gather*}
The built-in $i(A, S_1, S)$ inserts a number $A$ into the sorted list $S_1$ such that the resulting list $S$ is sorted.

\subsubsection{Runtime Repeated Recursion Unfolding}
Again we first have to find and define an appropriate rule template with sufficient simplification to improve on the runtime.

\paragraph{Unfolding}
Unfolding the recursive rule of $s/2$ results in the rule
\begin{gather*}
s(L, S) \simp L{=}[A,A_1|L_2] \, |\, 
              s(L_2, S_2), i(A_1, S_2, S_1), i(A, S_1, S). 
\end{gather*}
The number $A_1$ is inserted into the already sorted list $S_2$, then into the resulting list $S_1$, the number $A$ is inserted.
Repeating this unfolding scheme does not lead to any significant performance improvements, since we just generate more and more insertions.

\paragraph{Simplification}
The required simplification in this case is non-trivial.
In the above rule, we observe that we can more efficiently insert both numbers $A_1$ and $A$ during a single traversal of the list $S_2$.
We first insert the smaller number and then continue traversing the sorted list to insert the larger number. 
Since we get more and more insertions with each unfolding, we will actually have to insert more and more numbers in this way, and they have to be pre-sorted.
To implement this behavior, we use a built-in $m(S_1,S_2,S_3)$
instead of insertions.
It merges the sorted lists $S_1$ and $S_2$ into a sorted list $S_3$.

In the above rule, we first order $A$ and $A_1$ by putting them into a sorted list before they are merged with list $S_2$. For the necessary ordering we will also use $m/3$.
We replace the built-ins in the body of the rule
$i(A_1, S_2, S_1), i(A, S_1, S)$ by the semantically equivalent
$m([A], [A_1], S_0), m(S_0, S_2, S)$.
The simplified unfolded rule for sorting is now
\begin{gather*}
s(L, S) \simp L{=}[A,A_1|L_2] \, |\, 
              m([A], [A_1], S_0), s(L_2, S_2), m(S_0, S_2, S). 
\end{gather*}
The merging before the recursive call pre-sorts single numbers into a sorted list.
The merging after the recursive call merges this list into the sorted list returned by the recursive call.

Now let us unfold this simplified rule with itself. The resulting rule is
\begin{gather*}
s(L, S) \simp L{=}[A,A_1,A_2,A_3|L_3] \, |\, \\
              m([A], [A_1], S_0), m([A_2], [A_3], S_1),
              s(L_3, S_3), 
              m(S_1, S_3, S_2), m(S_0, S_2, S). 
\end{gather*}
We now generate more and more mergings.

\paragraph{Generalization}
Note that after the recursive call, we merge the list of two elements $S_1$ into the already sorted list $S_3$ and the resulting list $S_2$ is in turn merged with the two elements of the list $S_0$.
We can improve the runtime if we rearrange the mergings so that we merge lists that are about the same length.
We merge $S_1$ and $S_0$ first, move this merging before the recursive call and merge its result with $S_3$ after the recursive call:
\begin{gather*}
s(L,S) \simp L{=}[A,A_1,A_2,A_3|L_3] \, |\,\\
        m([A],[A_1],S_0), m([A_2],[A_3],S_1), m(S_1,S_0,S_4),
        s(L_3, S_3), 
        m(S_4, S_3, S). 
\end{gather*}
In this way we have almost halved the runtime by avoiding the generation and traversal of the intermediate sorted list $S_2$.

The introduction of mergings is the essential idea for the simplification of the unfolded rules. It gives rise to the rule template
\begin{gather*}
s(L,S) \simp L{=}[A,A_1,\ldots,A_m|L_1] \, | \, \mathit{Mergings}, s(L_1, S_1), m(S_0,S_1,S).
\end{gather*}
The placeholder $\mathit{Mergings}$ stands for the mergings of $A,A_1,\ldots,A_m$ that result in the sorted list $S_0$.

\subsubsection{Implementation}
We now implement the unfolding and the recursive constraint for sorting.
\paragraph{Unfolding with Simplification}
Relying on the rule template, 
the unfolding scheme is defined the following clause. 
\begin{verbatim}
simp_unf(
   Rule,                                         % given rule
   (s(L,S) <=> L=AL | MG, s(L2,S2), m(S0,S2,S))  % unfolded rule template
        ):-     
    copy_term(Rule, (s(L,S) <=> L=AL | MG1, s(L1,S1), m(S3,S1,S))), 
    copy_term(Rule, (s(L1,S1) <=> L1=AL1 | MG2, s(L2,S2), m(S4,S2,S1))),
    clean((MG1, MG2, m(S3,S4,S0)), MG),
    L1=AL1.
\end{verbatim}
We copy the input rule twice onto instances of the rule template to simulate the unfolding of the recursive call.
In the first copy, the recursive call is {\tt s(L1,S1)}. We directly use it as the head of the second copy of the given rule. 
The resulting unfolded rule is composed
of the head of the first copy {\tt s(L,S)},
of the guard of the first copy {\tt L=AL},
of the mergings {\tt MG1} and {\tt MG2} before the recursive call of the two copies together with {\tt m(S3,S4,S0)},
and the new merging after the recursive call {\tt m(S0,S2,S)}.
The built-in {\tt clean/2} removes superfluous {\tt true} constraints in the resulting mergings\footnote{The constraints stem from the original recursive clause and would proliferate otherwise.}.
Finally, the resulting guard is completed by executing the guard of the second copy {\tt L1=AL1} at unfolding time.
This will double the size of the open list {\tt AL} which ends in {\tt L1}.

\paragraph{Recursive Constraint}
For the sorting example, the {\tt rec\_unfold} rule is as follows:
\begin{verbatim}
rec_unfold @ sort(I,O) <=> 
    unf(s(I,O), [
       (s(A,E) <=> A=[C|B] | true, s(B,D), m([C],D,E)), 
       (s([],A)<=> true | A=[], true, true)
                ], URs), 
    mip(s(I,O), URs).
\end{verbatim}
We write the original recursive clause also in simplified form using {\tt merge/3} instead of {\tt insert/3}.

\paragraph{Unfolded Rules}
The first few rules that are returned by the unfolder with an appropriate query are
\begin{verbatim}
s(A,S) <=> A=[C,B,E,D,I,H,K,J|P] | 
         ((m([B],[C],G), m([D],[E],F), m(F,G,O)), 
          (m([H],[I],M), m([J],[K],L), m(L,M,N)), m(N,O,Q)), 
           s(P,R), m(Q,R,S).
s(A,K) <=> A=[C,B,E,D|H] | 
          (m([B],[C],G), m([D],[E],F), m(F,G,I)), 
           s(H,J), m(I,J,K).
s(A,G) <=> A=[C,B|D] | m([B],[C],E), s(D,F), m(E,F,G).
s(A,E) <=> A=[C|B] | true, s(B,D), m([C],D,E).
s([],A)<=> true | A=[], true, true.
\end{verbatim}
As with list reversal, the rule size roughly doubles with each unfolding, but again this does not increase the space complexity.

\subsubsection{Complexity}
We derive estimates for the time complexities.

\paragraph{Original Recursion}
The recursion depth is determined by the number of elements $n$ in the input list of the given query. In the original recursion we have $n$ recursive steps. 
In each step, {\tt insert/3} at worst traverses a list of length $n$.
This results in the well-known quadratic complexity $\Theta(n^2)$ of insertion sort. 

\paragraph{Unfolder} 
Copying head and guard of an unfolded rule and checking its guard is linear in the 
size of the open input list.
In {\tt simp\_unf/2} we basically copy the rule twice.
The runtime complexity $T_c(n)$ of a recursive step in the unfolder is therefore linear in the input list length $n$, $\Theta(n)$.

\paragraph{Meta-Interpreter}
Copying an unfolded rule and checking its guard is linear in the 
size of the open input list.
In the rule body, there are $n$ calls to {\tt merge/3} for an open input list of size $n$.
These mergings dominate the complexity.
The runtime of mergings is determined by the sum of the lengths of their input lists.
The mergings of the singleton lists involve the $n$ input list elements.
The mergings of the resulting two-element lists also involve all $n$ list elements.
The mergings of all lists of the same length always involve all $n$ input list elements.
The lists double their lengths until all elements are merged before the recursive call. So we have a number of different list lengths that is logarithmic in $n$. 

Overall, this results in a log-linear complexity for the mergings before the recursive call.
After the recursive call, a list of length $n$ is merged with the list resulting from the recursive call. The latter list cannot be larger than the former, because otherwise a more unfolded rule would have been applicable.
In conclusion, the runtime complexity $T_d(n)$ of a recursive step in the meta-interpreter is therefore log-linear in the input list length $n$, $\Theta(n \log(n))$.

\paragraph{Complexity of Runtime Repeated Recursion Unfolding}
The solution of the associated recurrence equation 
in accordance with Lemma \ref{complexities}
maintains the log-linear complexity 
$\Theta(n \log(n))$ in the input list length $n$ for the unfolder and the meta-interpreter together.
We therefore satisfy the sufficient and necessary condition for super-linear speedup according to Theorem \ref{speeduptheo}.
Note that the unfolder itself has a lower, linear complexity.
With repeated recursion unfolding the complexity is reduced from $\Theta(n^2)$ to $\Theta(n \log(n))$, clearly indicating a super-linear speedup.

\subsubsection{Benchmarks}

Table \ref{benchsort} shows benchmarks results for the sorting example.
Times are in seconds.
The benchmarks are performed with random permutations of integers from $1$ to $n$.
The individual runtimes show little variation, but are faster with already sorted lists, be they in ascending or descending order.
They confirm the super-linear speedup.

\paragraph{Original Recursion}
The experiments for the original version of insertion sort indicate a complexity that is indeed quadratic $\Theta(n^2)$.
Doubling the list length increases the runtime by a factor of four.

\paragraph{Unfolder and Meta-Interpreter}
The runtimes of the unfolder are consistent with a linear complexity $\Theta(n)$.
The meta-interpreter timings are consistent with a log-linear complexity $\Theta(n \log(n))$. 
The generation of all rules in the unfolder takes less time than applying one or more rules in the meta-interpreter.

\paragraph{Comparing Recursion Depths $2^i-1$ and $2^i$}
Going from input list length $2^i-1$ to $2^i$, 
the unfolder generates one more rule. 
It has twice the size of the previous rule.
And indeed the runtime for the unfolder almost doubles.
Going from list length $2^i-1$ to $2^i$, 
the meta-interpreter applies all unfolded rules in the first case but only the next more unfolded rule in the second case. 
In both cases, all rules are tried by checking their guard.
The runtime increases somewhat when going from $2^i-1$ to $2^i$.

\begin{table}[h] 
\begin{center}{
\begin{tabular}{l c l}

\begin{tabular}{|l|r|r|r|}
\cline{1-2}
\multicolumn{2}{|c|}{Original Sorting} \\
\cline{1-2}
Input $n$  & Time\\
\cline{1-2}
$2^{9}$ & 0.01 \\
$2^{10}$ & 0.03 \\
$2^{11}$ & 0.13 \\
$2^{12}$ & 0.51 \\
$2^{13}$ & 2.20 \\
$2^{14}$ & 8.61 \\
$2^{15}$ & 34.24 \\
\cline{1-2}
\end{tabular}

& &

\begin{tabular}{|l|r|r|r|}
\cline{1-4}
\multicolumn{4}{|c|}{Runtime Repeated Recursion Unfolding} \\
\cline{1-4}
Input $n$ & Unfolder & Interpreter & Total Time\\
\cline{1-4}
$2^{12}-1$ & 0.01 & 0.01 & 0.02 \\
$2^{13}-1$ & 0.01 & 0.02 & 0.03 \\
$2^{14}-1$ & 0.02 & 0.05 & 0.07 \\
$2^{15}-1$ & 0.04 & 0.11 & 0.15 \\
$2^{16}-1$ & 0.09 & 0.21 & 0.30 \\
$2^{17}-1$ & 0.19 & 0.47 & 0.66 \\
$2^{18}-1$ & 0.38 & 1.02 & 1.40 \\
$2^{19}-1$ & 0.77 & 2.24 & 3.01 \\
\cline{1-4}
$2^{12}$ & 0.01 & 0.01 & 0.02 \\
$2^{13}$ & 0.01 & 0.03 & 0.04 \\
$2^{14}$ & 0.04 & 0.06 & 0.10 \\
$2^{15}$ & 0.08 & 0.12 & 0.20 \\
$2^{16}$ & 0.16 & 0.27 & 0.43 \\
$2^{17}$ & 0.32 & 0.57 & 0.89 \\
$2^{18}$ & 0.65 & 1.34 & 1.99 \\
$2^{19}$ & 1.32 & 2.74 & 4.06 \\
\cline{1-4}
\cline{1-4}
\end{tabular}

\end{tabular}
\caption{Benchmarks for Sorting Example}
\label{benchsort}
}\end{center}
\end{table}

\section{Related Work}\label{related}

\emph{Program transformation} to improve efficiency is usually concerned with a strategy for combining unfolding and folding to replace code
(for an overview see e.g. \cite{pettorossi1996rules,visser2005survey,pettorossi2024historical}). 
The transformations are typically performed offline, at compile-time.
{Program transformation} for specific aims and 
applications is abundant in logic programming in general \cite{pettorossi1999synthesis} and in CHR in particular
\cite{chr_survey_tplp10,Fruehwirth15}. 
General methods exist for
unfolding \cite{gabbrielli2015unfolding} (which we have adapted for this paper), for specializing rules with regard to a specific given query \cite{fru_specialization_lopstr04}, 
and for optimizations induced by confluence \cite{abd_fru_integration_lopstr03}.
More recently, \cite{10.3233/FI-2017-1461} uses program transformation implemented in CHR on constraint logic programs that verify properties of imperative programs.

{\em Partial evaluation} is a program transformation to execute programs with partially known input to specialize it, typically at compile-time.
Simple partial evaluation alone cannot achieve super-linear speedup (Chapter 6 in \cite{Jones-etal93}). This linear speedup is called Type 1 speedup in \cite{Jones04}. This result does not apply to our approach, because we strongly rely on rule simplification.
Involving an interpreter, our approach belongs to Type 2 speedup according to \cite{Jones04}.
\emph{Polyvariant program specialization} is the generation of specialized versions of a program according to different constraints that restrict its execution \cite{angelis_2022,gallagher2019polyvariant}. 
One could argue that repeated recursion unfolding shares the same underlying idea, since it generates versions of a single rule specialized by recursion length.

In general, {\em super-linear speedups} by program transformation are rare and mostly concern parallel programs.
Our technique applies to sequential programs.
In a sequential setting, 
super-linear speedups can sometimes be achieved with {\em memoization}, where the results of recursive calls are cached and reused if the same recursive call reappears later on. A classical example where this runtime optimization applies is the naive double recursive implementation of the Fibonacci function. Typically, memoization pays off with multiple recursion, while our approach at the moment works with linear recursion.
{\em Tupling} \cite{hu1997} is another technique that can achieve super-linear speedup. It applies when several recursions operate on the same data structure. Then tupling tries to merge these recursions into a single one. 
Memoization and tupling can be regarded as special cases of folding.
Then there is work based on \emph{supercompilation} for functional programming languages like Refal and Haskell. 
In advanced cases of this offline program transformation such as distillation \cite{hamilton09} and equality indices \cite{glueck16},
sophisticated generalization while unfolding increases the chance for folding
and can achieve super-linear speedup on some examples.
In contrast, our approach so far does not involve folding
and works online at runtime.
It requires a problem-specific simplification that has to be provided at compile-time.

The notion of {\em repeated recursion unfolding} was introduced in \cite{reclopstr}.
But there the rules were transformed at compile time. Because of this, super-linear speedup was only possible for calls that did not exceed a given fixed number of recursion steps.
For larger calls, the speedup detoriated to a constant factor. 
Here we substantially revised and greatly extended the approach for just-in-time (JIT) online execution.
We introduced an unfolding scheme and a specialized meta-interpreter so that super-linear speedup can be made possible on-the-fly at runtime for any recursive call.
Our technique relies solely on unfolding and simplifying the recursive step again and again. It ignores the base case of the recursion.
We add redundant rules this way but never remove any. 
We never fold a recursive rule, but we simplify rule bodies (recursive steps).

As pointed out by a helpful reviewer,
the basic idea of {\em repeatedly unfolding} a structural recursion is sketched in work \cite{Millroth92,Millroth95} for Reform Prolog.
In Prolog, there are no guards, so unfolding is rather straightforward.
The unfolding in the cited work serves a different purpose, to parallelize the recursive steps.
AND-parallelism requires that the recursive steps are (made) rather independent of each other, while in our approach it is essential to simplify the recursive steps for a sequential computation. This simplification benefits from dependence between the recursive steps. In this sense, the approach of Reform Prolog and of runtime repeated recursion unfolding are complementary: where simplification is not sufficient to achieve super-linear speedup, parallelization could be considered.

Compile-time {\em recursion unrolling} \cite{Rugina:2000:RUD:645678.663942} for C inlines (unfolds) recursive calls, fuses (merges) conditionals and then re-rolls (folds) back the recursive part of the procedure to ensure a large simplified base case. 
The transformation is repeatedly applied, each time increasing the recursion depth by one.
This technique is presented for double recursive divide and conquer algorithms where it can result in a constant factor speedup. 
In our approach, we work at runtime with linear recursion. We do not touch the base case at all and we do not fold. Repeated unfolding in our approach results in a doubling of the recursion depth covered, while in recursion unrolling, recursion depth is increased by one only. Conditional fusion merges only identical conditions, typically from the base case, while in our approach arbitrary guards are merged during unfolding. Recursion unrolling is cited mainly in work for parallelization and hardware programs, while we aim at sequential computations in software.

Recursion unrolling is derived from
{\em loop unrolling} (Chapter 10.4.5 in \cite{aho2006compilers},
\cite{leopoldseder2018})
which is a standard code transformation in compilers that repeats the body of a loop a small fixed number of times. In this way, the overhead of the resulting code and the number of loop iterations can be reduced. On the other hand, prologue and epilogue code has to be added to account for the cases where the number of loop iterations is not a multiple of the iterations covered by the unrolled loop. The resulting speedup is a constant factor, typically less than two.

Finally, runtime repeated recursion unfolding should not be confused with {\em recursive doubling} \cite{kogge1973,afrati2012} (also called binary splitting in mathematics). In this optimization method, a problem is split top-down into two separate sub-problems of equal complexity that can be executed in parallel. Hence, a linear recursion would be transformed into a double recursion following the divide and conquer approach.
In our method, we merge subsequent recursive steps and simplify them, doubling the number of recursive steps covered with each unfolding. No double recursion is introduced.

\section{Discussion}\label{discuss}

We discuss some issues and limitations of runtime repeated recursion unfolding and suggest some possible improvements as well.

\myparagraph{Rule Simplification}
Our technique hinges on sufficient unfolding simplification of the recursive step resulting from unfolding. This simplification has to be provided at compile-time. It requires some insight into the given problem and cannot be fully automated (but mathematical software tools and theorem provers might help).
Any existing optimization technique can be applied such as all kinds of program transformation.
If the recursive steps are {\em arithmetic computations with polynomials}, they could be optimized using efficiently computable representations such as Horner's method or more advanced approaches such as \cite{leiserson2010}. Another possibility is to use results from the verification of loops to compute closed forms (loop summarization \cite{Kincaid2019}, loop acceleration \cite{frohn2020}, loop solving \cite{Kafle2021}) for sequences of recursive steps.
While promising for arithmetic computations, this approach does not apply to structural recursion.

Often, simplification relies on {\em algebraic properties of the operations} in the recursive step.
Judging from our experiments, we see that all examples involve the use of associativity of the operations to regroup the computation so that it becomes more efficient. 
For summation, it is the associativity of addition, for list reversal that of list concatenation, for sorting that of ordered merging.
Another commonality is the standard optimization technique of finding {\em common sub-expressions}, i.e. to merge repeated data or operations.
For summation, $N{+}N$ is replaced by $2{*}N$, for reversal and sorting, repeated traversals of lists are (partially) merged.
To achieve this, we may need to replace operations, e.g. addition by multiplication for the summation example and insertion by merging for sorting.

Clearly, an algorithm implementation that is already optimal cannot be further improved. For a simple example, a search for the minimum of an unordered list has to go through all elements of the list. We cannot improve the time complexity of the linear direct recursion that performs this traversal without changing the data structure.
A algorithm that keeps intermediate results of recursive steps can also be hard to optimize. 
For a simple example, this applies to a recursion that squares each number in a given list. But if the list contains successive integers, we can optimize the computations.

As one reviewer remarked, the necessary simplification scheme requires some effort, so one could go all the way deriving a more efficient algorithm for the recursion at hand. In particular, one could say our sort example is halfway towards deriving merge sort from insertion sort. On the other hand, the simpliciation in our list reversal example is not related to the common efficient version of reversal using an additional accumulator argument. 
The strength of our approach is that it provides a systematic practical and formally proven correct way to explore possible speedups and gives theorems when a super-linear speedup can be achieved.

\myparagraph{Limited Recursion}
Our approach as presented is restricted to single recursive rules.
This does mean a loss of generality in terms of expressiveness.
Any kind of recursion can be expressed as a linear recursion using continuation passing (such as in the rule-based language BinProlog \cite{tarau2012}). The resulting linear recursive rules can be merged into a single such rule by introducing an auxiliary constraint performing the recursive steps.
However, our preliminary experiments indicate that the resulting single linear recursive rule may be awkward and hard to optimize. 
Thus future work should consider multiple recursive rules directly.
When insisting on the optimal rule application strategy, a naive extension of our approach could lead to a combinatorial explosion in the number of unfoldings.

\myparagraph{Limited Time Complexity}
We have considered complexity classes of the form 
$n^j \log(n)^k$, where the parameter $n$ is the recursion depth.
This allowed us to prove precise tight complexity results.
However, complexity is usually stated in the size of the given problem. 
The size $s$ often coincides with the recursion depth $n$ (as was the case in our benchmarked problems), but it must not.
For example, finding an element in a binary search tree by recursion has a complexity linear in the depth of the tree, but logarithmic in the size of the tree.
This poses no problem, as
our results carry over to complexity parametrized by problem size. To change the parameter,
it suffices to find a non-constant positive monotonic function from size $s$ to recursion depth $n$ and replace $n$ with it in the complexities.
For the tree search, this gives $n{=}\log_2(s)$ and replacing $n$ by $\log_2(s)$ leads to the correct complexity results.
Exponential complexity in terms of size $s$ is also possible and in this way a super-linear speedup into a polynomial complexity can be modeled.
For upper bounds on complexity, it suffices that the function limits $n$ from above.
Therefore our focus on tractable problems in terms of recursion depth does not preclude arbitrary complexities expressed in terms of problem size.
In particular, our results also apply to exponential problems.

\myparagraph{Space Complexity}
Another issue are the space requirements of our approach.
We generate a number of rules that is logarithmic in the recursion depth of the given query.
In our examples we saw an increase in \emph{rule size}. 
With each unfolding, the rule size roughly doubled. 
In effect, the size of all unfolded rules taken together is proportional to 
the size of the query, i.e. input number for the summation example and
to length of the input list for reversal and sorting.
Hence there was no increase space complexity.
In general however, we cannot rule out code explosion in our approach.

\myparagraph{Limited Unfolding}
Rule unfolding in CHR has some conditions and may not be possible at all.
Second, repeated unfolding may not produce enough rules to allow for optimal rule applications.
So far, we have not observed these problems in practice. If they should occur, then we think they could be tackled with a more liberal definition of unfolding in CHR.

\myparagraph{Possible Improvements}
Note that unfolded rules are generic and can be reused for any later call, improving the efficiency further.
As for the implementation, the following optimizations come to mind:
The unfolder and the meta-interpreter can be specialized for a given recursive rule using standard partial evaluation techniques, which typically lead to an additional constant factor speedup.
The unfolder and the meta-interpreter are currently head-recursive, the implementation could be made tail-recursive.
Finally, one reviewer suggested that one could generalize the approach.
Instead of increasing the number of recursive steps by a factor of 2 during unfolding, once could use other factors. With larger factors, we need less unfolded rules, but may have to apply some of them several times. In some cases, this may lead to a runtime improvement. However, our worst-case time complexity results would not be improved, because the runtime can only decrease by a constant factor at most in this way.

\section{Conclusions and Future Work}

We have introduced a strategy for online program optimization that is based on existing techniques such as unfolding that can achieve super-linear speedup.
We have given a formal definition of runtime repeated recursion unfolding with simplification and proven its correctness.
Our technique generates several versions of a single linear direct recursive rule for a recursive call at runtime, where each version doubles the number of recursive steps covered. The base case of the recursion is ignored.
Our just-in-time method reduces the number of recursive rule applications to its logarithm at the cost of introducing a logarithmic number of unfolded rules.
We provided a lean implementation of our approach in five rules, comprising the unfolder and the meta-interpreter and analyzed its complexity using recurrences.
In our speedup analysis, we proved a sufficient condition as well as a sufficient and necessary condition for super-linear speedup relating the complexity of the recursive steps of the original rule and the unfolded rules.
The results rely on an optimal rule application strategy that we proved sound and complete.

We showed with benchmarks on three simple basic algorithms that the super-linear speedup indeed holds in practice.
For each example, we had to find a specific rule unfolding scheme.
For ease of implementation, we used rule templates.
Table \ref{speedupexamples} summarizes our estimated and observed time complexity results for our examples. They feature typical complexities of tractable algorithms and reduce the time complexity by a factor of $\Theta(n)$ or $\Theta(n / \log(n))$. 
For list reversal, the complexity of the given recursion was reduced to that of its recursive step.
Summation and list reversal are examples for satisfying the sufficient condition for super-linear speedup.
The sorting example does not, but satisfies the sufficient and necessary condition, with different complexities for the unfolder and meta-interpreter.
\begin{table}
\begin{center}{
\begin{tabular}{|l|c|c|c|c|}
\cline{1-5}
Example & R.Step $T_b(n)$ & Recursion $T_r(n)$ & R.Step $T_c(n){+}T_d(n)$ & Unfolded $T_u(n)$\\
\cline{1-5}
\cline{1-5}
Summation & $\Theta(\log(n))$ & $\Theta(n \log(n))$ & $\Theta(\log(n)+\log(n))$ & $\Theta(\log(n)^2)$\\
\cline{1-5}
List Reversal & $\Theta(n)$ & $\Theta(n^2)$ & $\Theta(n+n)$ & $\Theta(n)$\\
\cline{1-5}
Sorting & $\Theta(n)$ & $\Theta(n^2)$ & $\Theta(n + n \log(n))$ & $\Theta(n \log(n))$\\
\cline{1-5}
\end{tabular}
\caption{Summary: Time Complexity Classes of Super-linear Speedup for Examples}
\label{speedupexamples}
}\end{center}
\end{table}

Overall, runtime repeated recursion unfolding provides a general strategy for online optimization of linear direct recursions in which the sufficient simplification of successive recursive steps leads to predictable speedups.

\myparagraph{Future work}
This paper introduces our approach, but does not explore it in full.
Our main limitation is the challenge of finding sufficient problem-specific simplifications.
Future work should investigate classes of functions that can be simplified in the necessary way, such as polynomial arithmetic expressions.
We assumed a single recursive rule with linear direct recursion written in CHR.
As we have discussed, 
this does not result in a loss of generality in terms of expressiveness.
However, this restriction may lead to unnatural implementations that are hard to optimize. 
Hence 
we want to extend our technique to mutual and multiple recursion as well as multiple recursive rules \cite{recunfoldprolog}.

We defined and implemented repeated recursion unfolding using the rule-based language CHR, but we think our approach can be applied to other rule-based languages and mainstream programming languages as well.
First candidates are other declarative programming languages like Prolog and Haskell.
In Prolog we will have to deal with non-determinism in the rule choice, in functional languages we will have the issue of nested guards and conditionals.
For the implementation, meta-programming features may not be necessary if the interpreter is specialized with regard to the given recursion so that the meta-calls go away.
It already might be an advantage that the number of recursive steps is reduced to its logarithm by our approach.
For example, there is a limit on recursion depth in languages like Java and Python due to the limit on stack size.
Last but not least, it should also be possible to apply our technique to loops instead of recursion.

\medskip
{\bf Acknowledgements.} 
This research work was initiated during the sabbatical of the author in the summer semester of 2020.
We thank the anonymous reviewers and Sascha Rechenberger for comments.
In particular, one reviewer made highly detailed and substantial comments and suggestions that helped tremendously to clarify, improve and extend the paper.

\bibliographystyle{fundam} 
\bibliography{recunfold,biblio,chrjust}

\end{document}